\documentclass[final,3p,11pt]{elsarticle}

\biboptions{sort&compress}   
\usepackage{amsmath}         
\usepackage{hyperref}        
\hypersetup{colorlinks=true, linkcolor=blue, citecolor=blue, urlcolor=blue}
\usepackage[capitalize,nameinlink,noabbrev]{cleveref} 
\usepackage{tabularray}     
\UseTblrLibrary{booktabs}   
\usepackage[labelformat=simple]{subcaption}      
\usepackage[section]{placeins}   
\usepackage{array}               
\usepackage[autostyle]{csquotes} 
\usepackage{pdflscape}           
\usepackage{bm}                  
\usepackage{caption}             

\journal{}
\graphicspath{{figures/}} 

\begin{document}

\begin{frontmatter}

\title{Development of a Methodology for the Automated Spatial Mapping of Heterogeneous Elastoplastic Properties of Welded Joints}

\author[label1,label2]{Robert Hamill\corref{cor1}} 
\cortext[cor1]{Corresponding author}
\ead{robert.hamill@ukaea.uk}
\author[label2]{Allan Harte} 
\author[label1]{Aleksander Marek} 
\author[label1,label3]{Fabrice Pierron} 

\affiliation[label1]{organization={Faculty of Engineering and Physical Sciences, University of Southampton},
addressline={Highfield}, 
city={Southampton},
postcode={SO17 1BJ}, 
state={Hampshire},
country={UK}}

\affiliation[label2]{organization={United Kingdom Atomic Energy Authority, Culham Centre for Fusion Energy},
addressline={Culham Science Centre}, 
city={Abingdon},
postcode={OX14 3DB}, 
state={Oxfordshire},
country={UK}}

\affiliation[label3]{organization={MatchID NV},
addressline={Leiekaai 25A}, 
city={Ghent},
postcode={BE 0563. 512.491}, 
state={East Flanders},
country={Belgium}}

\begin{abstract}
Knowledge of the mechanical properties of materials is required for the design and analysis of engineering products, however, the characterisation of heterogeneous properties using traditional techniques is limited by spatial resolution or insufficient reliability. This paper presents a novel methodology for the characterisation of heterogeneous mechanical properties by extending the virtual fields method through the automated spatial parameterisation of constitutive parameters. Collaboration with the United Kingdom Atomic Energy Authority provided this project with an application focus on the characterisation of the spatially-varying, elastoplastic mechanical properties of welded joints. The developed methodology enables the novel characterisation of welds with assorted geometries, varied loading configurations and dissimilar materials. Numerical verification of the developed method was performed using synthetic data equivalent to that obtained experimentally using optical measurements, where the kinematic fields are known and controlled. The results confirm that the proposed approach converges towards the target parameter maps without any \textit{a priori} information on the distribution of the properties, successfully demonstrating the established methodology as a proof of concept.

\end{abstract}

\begin{graphicalabstract}
\includegraphics{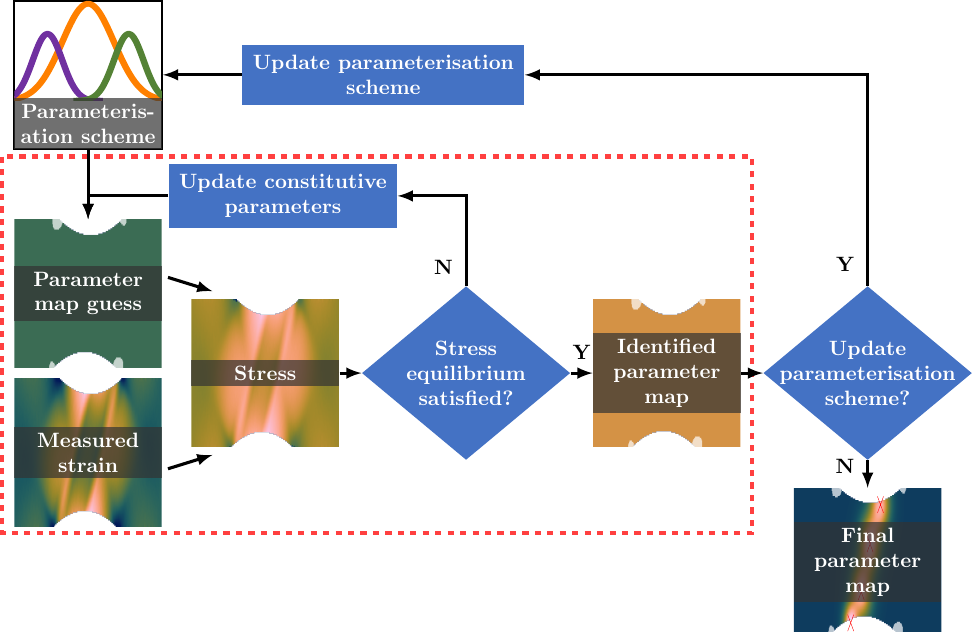}
\end{graphicalabstract}

\begin{highlights}
\item Welded joints remain a critical structural integrity challenge in many industries
\item Traditional techniques cannot reliably resolve spatially varying mechanical properties
\item A novel methodology maps the heterogeneous elastoplastic properties in welds
\item Spatially resolved data could reduce conservatism and accelerate qualification
\end{highlights}

\begin{keyword}
virtual fields method \sep inverse identification \sep characterisation \sep welding \sep materials testing 2.0 \sep heterogeneity
\end{keyword}

\end{frontmatter}

\newcommand{\stress}{\sigma}
\newcommand{\bodyforce}{b}
\newcommand{\density}{\rho}
\newcommand{\accel}{a}
\newcommand{\vdisp}{u}
\newcommand{\strain}{\varepsilon}
\newcommand{\traction}{T}
\newcommand{\normal}{n}
\newcommand{\vol}{V}
\newcommand{\surf}{S}
\newcommand{\surfTraction}{S_f}
\newcommand{\lineTraction}{L_f}
\newcommand{\thickness}{h}
\newcommand{\areaPoint}{s}
\newcommand{\numPoints}{N_p}
\newcommand{\resForceVec}{\bm{F}}
\newcommand{\resForce}{F}
\newcommand{\appliedForce}{F_A}
\newcommand{\reconForce}{F_R}
\newcommand{\IVW}{\mathrm{IVW}}
\newcommand{\EVW}{\mathrm{EVW}}
\newcommand{\FRE}{\mathrm{FRE}}
\newcommand{\EGI}{\mathrm{EGI}}
\newcommand{\wgt}{\alpha}
\newcommand{\sbvfCost}{\phi_{\mathrm{SBVF}}}
\newcommand{\numVF}{N_f}
\newcommand{\sliceWidth}{w}
\newcommand{\sliceLength}{L}
\newcommand{\numSlices}{N_r}
\newcommand{\numSteps}{N_t}
\newcommand{\numWin}{N_m}
\newcommand{\numParam}{N_K}
\newcommand{\constitParam}{K}
\newcommand{\interp}{\phi}
\newcommand{\kernel}{B}
\newcommand{\rbfWeight}{\lambda}
\newcommand{\rbfCentre}{c}
\newcommand{\covMat}{\Sigma}
\newcommand{\rotAngle}{\theta}
\newcommand{\numBasis}{N_b}
\newcommand{\rbfVarOne}{\sigma_1^2}
\newcommand{\rbfVarTwo}{\sigma_2^2}
\newcommand{\mixWeight}{\lambda}
\newcommand{\globalCost}{\phi}
\newcommand{\egiCost}{\phi_{\mathrm{EGI}}}
\newcommand{\freCost}{\phi_{\mathrm{FRE}}}
\newcommand{\numWinSizes}{N_k}
\newcommand{\winWeight}{\gamma}
\newcommand{\winLength}{L}
\newcommand{\egiScale}{a}
\newcommand{\freScale}{b}
\newcommand{\stressRef}{\stress_{\mathrm{ref}}}

\section{Introduction}\label{sec:intro}
Traditionally, mechanical property characterisation is conducted using homogeneous specimens under simple loading for which strain and stress distributions can be readily assumed. Such approaches only provide the averaged behaviour of heterogeneous specimens, however, knowledge of the spatially varying properties would enable safer, more efficient design and analysis of engineering products. Heterogeneity can be introduced intentionally, in composite materials for example, or it may arise as a consequence of a component's manufacturing process or service life. Collaboration with the United Kingdom Atomic Energy Authority provides this work with an application focus on the characterisation of welded joints, although the proposed methodology will be applicable to other instances of heterogeneity. In this work, the terms homogeneous and heterogeneous are used to describe the distribution of mechanical properties throughout a material sample. Furthermore, the length-scale of interest is that of the bulk material rather than the microstructure.

\section{Background}\label{sec:background}
\subsection{Heterogeneity in welded joints}
The design and performance validation of welded joints remains a critical structural integrity issue in many engineering industries, including nuclear fusion. Welds are often sites of failure because they combine geometric discontinuity with strong gradients in microstructure, mechanical properties and residual stress. Furthermore, weld performance is typically sensitive to manufacturing variability and the constraint state in the as-built component. Fabrication rules typically necessitate non-destructive testing and prescribe allowable tolerances for defects such as porosity, cracking, and geometric discontinuity. Nevertheless, qualification approaches must consider the remaining variability.  For example, design codes such as RCC-MRx require representative coupons to support a test matrix covering multiple mechanical loading modes and, where relevant, environmental conditions, \textit{e.g.} temperature. In fusion applications, performance under complex conditions (such as irradiation and coolant interaction) must be demonstrated to satisfy integrity assurance in first-of-a-kind environments, with sufficient repeat specimens to capture variability. While these codes use trusted standards to generate material properties from a test matrix, the representativeness of the coupons used is uncertain, because full scale structures commonly differ in thickness, constraint, residual stress state and local heterogeneity. 

This gap between laboratory scale conditions and structural component conditions is well recognised in the structural integrity community; the typical response is to mitigate risk by increasing conservatism in design rules, inspection frequency and lifetime prediction. In the fusion structural integrity community, elasto-plastic design-by-analysis approaches are under development to reduce conservatism in design for greater functionality (\textit{e.g.} thinner sections for reduced tritium inventory) \citep{wang2025ens,degiorgio2025rcc}. A key limitation in these efforts is the lack of spatially resolved property data, as traditional characterisation techniques generally only provide averaged responses.  Knowledge of heterogeneous mechanical properties would improve the fidelity of structural analyses, enhancing insight into failure mechanisms and reducing conservatism, while also informing manufacturing processes such as welding and post-weld heat treatment.  In turn, this provides a pathway to accelerate qualification: instead of relying on large test matrices and repeated tests to define uncertainty bounds, fewer information-rich tests can be used to quantify the spatial variability that governs performance. This is particularly important for nuclear fusion power plant delivery, where the time scale for qualification of novel materials requires an accelerated approach. 

\subsection{Existing methods for the characterisation of welded joints}
\paragraph{Welding simulation}
Some authors have proposed using numerical simulation to estimate a weld's resulting mechanical properties \citep{zuniga1995det,bleck2010met,wei2015exp,nazemi2019str}. For example, \citeauthor{nazemi2019str} developed a fully coupled multi-physics simulation accounting for thermal, metallurgical and mechanical aspects of welding \citep{nazemi2019str}. Alternatively, \citeauthor{wudtke2015hie} proposed a homogenisation approach for estimating the mechanical properties of a welded joint using representative volume elements (RVEs) \citep{wudtke2015hie}. These RVEs are defined using microstructural imaging, and then assigned mechanical properties from literature. Although these methods have the potential to yield accurate results -- they require complex simulation and there is still the need for mechanical characterisation of the as-welded specimens to validate the simulations. 

\paragraph{Miniaturised tensile testing (MTT)}
MTT aims to extract and test homogeneous specimens from a component with regular heterogeneity. For example, miniature specimens could be extracted from the weld metal, heat affected zone (HAZ) and parent metal of a welded joint. \citeauthor{cam1998inv} and \citeauthor{lavan1999mic} helped popularise the MTT method for welded joints through their work demonstrating the characterisation of mechanical properties across electron beam welded  joints \citep{cam1998inv,lavan1999mic}. Much work has since been published characterising different materials and weld configurations, however, they often share common features \citep{lavan2001loc,molak2009mea,rao2010cha,pakdil2011mic,wang2013loc,hertele2014adv}. 

Typically, microstructural imaging is used to identify various regions of the welded joint, then electrical discharge machining is used to extract specimens. However, if the specimen's dimensions fall below a critical length scale, its properties will no longer reflect the bulk properties of the region it has been extracted from (a phenomenon referred to as \enquote*{specimen size effect} \citep{zheng2020sta}). For the testing itself, specialised rigs are generally used to minimise gripping and alignment issues, and more recently the deformation is measured using optical methods. Due to the increased relative effects per unit volume, miniature specimens are highly sensitive to surface imperfections, residual stresses and experimental setup errors. As such, there is typically greater variation between tests, and multiple tests are required to ensure reliability of the results. In addition, extracting multiple representative specimens from each region may not be possible. In 2020, \citeauthor{zheng2020sta} published a review of MTT which can be referred to for further details \citep{zheng2020sta}.

\paragraph{Indentation hardness testing}
Mechanical characterisation using hardness testing typically uses instrumented indentation in which the depth of penetration and applied load are continuously recorded throughout the test. Recent work can be grouped two main approaches: (i) direct or representative stress-strain methods, and (ii) inverse finite element approaches.

The direct approach seeks to develop empirical relationships between the load-displacement curve and tensile properties. To avoid the need for imaging, the contact area can be estimated using an area function dependent on the measured contact depth and the indenter shape - assuming sink-in occurs at the contact periphery. This contact area is then used in combination with the load-displacement curve to calculate tensile properties \citep{oliver1992imp,oliver2004mea}. For elastic properties only the unloading portion of the curve, which is assumed to be entirely elastic, is required. Obtaining elastoplastic properties is more challenging, but typically the load-depth curve is related to a selected constitutive model by means of dimensionless equations which are empirically fitted using experimental or finite element data. Many such algorithms have been proposed using a range of constitutive models and indenter shapes \citep{dao2001com,ye2013use,evdokimov2018mec}. Once established, the empirical relationships require no costly simulation and can be quickly used to obtain results. However, the generation of these functions requires a significant amount of data to ensure robust correlation of the load-depth curves to accurate numerical results. The relations are also difficult to generalise for multiple materials without introducing large errors or sensitivities. For example, functions developed for the characterisation of a particular grade of steel would provide erroneous results for an alternative grade. 

The second approach utilises iterative finite element simulations of the indentation test to obtain a best-fit set of constitutive parameters that minimises a difference between simulated and actual data. Typically, load-displacement values or the residual indent profiles are used \citep{campbell2021cri}. This approach has been used to obtain the elastoplastic properties of electron beam and laser beam welded joints \citep{debono2017rob,javaheri2020mec}. The model must be sufficiently representative of the indentation experiment, which requires knowledge of complex boundary conditions and interfacial friction.

In general, indentation hardness testing is relatively convenient to perform, non-destructive, moderate spatial resolution and widely used. However, many uncertainties still surround the use of indentation for mechanical property identification. The properties obtained using indentation are averaged quantities characteristic of a length-scale defined by the indentation depth or indenter's radius \citep{debono2017rob}. If this length-scale is too small, the resulting indentation data will be representative of an individual phase or grain, rather than the bulk. As the plastic response of a material is governed by microstructural features, the region being plastically deformed must be large enough to reflect that of the bulk. Another key issue is that of uniqueness. If multiple sets of constitutive parameters produce the same penetration load-depth data, it is not feasible to uniquely identify the material properties of the specimen. This is a well-known issue, with some authors proposing the use of multiple indenter geometries. However, \citeauthor{chen2007uni} found this did not always solve the issue of uniqueness \citep{chen2007uni}. In any case, more fundamentally, many factors affect the results of indentation testing and uni-directional indentation at the surface of the material provides limited data. Indentation testing can be extremely useful for qualitative investigation of property variation across a welded joint. It can also be beneficial for validating particular properties once the procedure has been well calibrated. However, currently, the method lacks robustness for the general application of characterising heterogeneous mechanical properties of a welded joint. 

\paragraph{Inverse identification methods}
Inverse identification strategies have been developed to take advantage of the information rich, full-field deformation measurements generated using non-contact optical techniques, such as digital image correlation (DIC). A comparison of popular inverse methods can be found in \citep{avril2008ove,martins2018com,pierron2023mat}. For the mechanical property identification of welded joints, the most relevant methods are: the uniform stress method (USM), finite element model updating (FEMU) and the virtual fields method (VFM). 

The USM assumes each cross-section of the specimen has uniform uniaxial stress, equal to the applied load divided by the cross-sectional area. This approach is therefore limited to uniaxial loading of butt welds for which the variation of the weld properties can be considered 1D. Strains computed from the full-field deformation are plotted against the approximated stresses, and a constitutive model fitted to the stress-stain data. The USM designation was first used to describe the approach in \citep{sutton2008ide}. 
FEMU adjusts a finite element model of the experiment until the residual between experimental and numerical results is minimised. The data used for comparison does not have to be full-field, however, full-field displacement or strain data are typically used for mechanical parameter identification. 
The VFM combines the measured strain fields with a postulated constitutive model and initial guess of parameters to reconstruct the stress fields. The static admissibility of this constructed stress field is then checked using the principle of virtual work (PVW), which assesses equilibrium between the reconstructed stress field and the external forces. The identified parameters are determined to be those that minimise the residual of the PVW. The theory of the VFM is discussed further in \cref{sec:stress_equil_metrics}.

In a uniaxial tensile test, the integral sum of the longitudinal stress across the specimen cross-section must equal the applied force at any position along the specimen length. However, the actual distribution of longitudinal stress may not be uniform. Unlike the USM, the VFM and FEMU do not assume a particular stress distribution -- making them better suited for experiments with more complex geometries or boundary conditions. 
The primary advantage of the VFM versus FEMU is its computational efficiency as no resolution of the direct problem is required (\textit{i.e.} no calculation of strain from the boundary conditions is required). The VFM is also less restricted by full knowledge of boundary conditions, as although resultant forces are required, the distribution of tractions is not. Contrastingly, FEMU is inherently iterative and requires accurate modelling of the experimental geometry and boundary conditions. A disadvantage of the VFM is that it requires an assumption of the through-thickness stress distribution as generally only surface measurements are known. FEMU is therefore more suited for complex 3D specimens. In the case of heterogeneous specimens, for which more parameters need to be identified, the computational efficiency of the VFM is significant.  

In 1999, \citeauthor{reynolds1999dig} used the USM to identify the yield strength and strain hardening exponent for three discrete regions of a friction stir welded aluminium 5454 specimen \citep{reynolds1999dig}. The regions were defined \textit{a priori} using microstructural examination. Later work examined the validity of the uniform stress assumption by applying the USM to a friction stir welded aluminium 2024 specimen \citep{lockwood2003sim}. In 2006, \citeauthor{boyce2006con} applied the method to laser beam welded, stainless steel specimens with a fusion zone width of only 0.65~mm \citep{boyce2006con}. These works demonstrate that the USM is adequate for simple specimen geometries under uniaxial loading, however, each author highlighted that the approach is limited by the strength of the weakest region. If one region is weaker, that region will likely fail before the entire constitutive relationships of the other regions can be identified. This limitation can be circumvented through experiment and specimen design, particularly when the uniform stress condition no longer has to be assumed. 

\citeauthor{sutton2008ide} used both the USM and the VFM to identify the elastoplastic properties throughout an arc welded joint in grade 690 steel \citep{sutton2008ide}. The region of interest was manually discretised into seven predetermined zones using microstructure and preliminary testing. Each region was assumed locally homogeneous with the elastic limit, hardening modulus and hardening exponent remaining spatially constant throughout each region, but differing between regions. Young's modulus was assumed to be constant throughout all regions. The USM was used in the same way as before. For the VFM, the stress fields were reconstructed using the measured strains, and a Johnson-Cook plasticity model with the piecewise parameters assigned to each region. The virtual fields were defined to assess stress equilibrium in a slice-wise manner along the length of the specimen. The two methods produced similar results due to the relatively small widths of the specimen and the uniaxial loading condition. The authors noted that \enquote*{similar computations using a finite element updating based approach would last more than one hundred times longer} \citep{sutton2008ide}.

\citeauthor{lelouedec2013ide} developed on \citeauthor{sutton2008ide}'s approach by using the same slice-wise discretisation for both the evaluation of stress equilibrium and the parameterisation of the constitutive parameters \citep{lelouedec2013ide}. The aluminium 5456 friction stir welded specimen was discretised into fifty-three slices, each with a width of one strain data point. Hence, this discretisation assumes the weld is homogeneous in the width direction and properties only vary along the length. A linear isotropic hardening model was used, with the yield stress and hardening modulus being identified for all fifty-three slices in under eight minutes.

\citeauthor{saranath2014zon} applied the USM and the slice-wise VFM approach to mild steel, butt-welded specimens manufactured using GMAW and laser beam welded titanium alloy specimens \citep{saranath2014zon,saranath2015loc}. For the former, they discretised the weld according to the microstructure and strain measurements. In the latter, they used the same slice-wise approach discussed above. \citeauthor{saranath2015loc} concluded by stating the  \enquote*{VFM is highly recommended for zone wise local characterization of welds}. 

\citeauthor{rossi2020stu} used the Fourier series implementation of the VFM (F-VFM) to identify the spatial distribution of the hardening behaviour of tailor heat treated blanks \citep{rossi2020stu}. In the F-VFM, the virtual fields are expressed as combinations of sine and cosine functions with various spatial frequencies, and the corresponding Fourier coefficients represent the distribution of properties \citep{nguyen2014fou}. A Voce hardening law was used, with the one-dimensional variation of a hardening parameter identified along the specimen's length. 
 
\citeauthor{andrade-campos2020int} used FEMU to characterise an aluminium 6082 friction stir welded joint \citep{andrade-campos2020int}. On the basis of indentation hardness results, the weld was discretised into four regions and approximated as symmetric. 

Many of the papers published characterising various welded joint materials use one of the methodologies discussed above \citep{kartal2007det,acar2009var,leitao2012det,peng2018det, bai2018vir,nazemi2019str,shahmirzaloo2021eva,wu2020det,ramachandran2021app,lattanzi2021fas}. 

\paragraph{The need for automated parameterisation}
MTT is an effective technique if representative homogeneous specimens can be extracted from a predetermined region. However, specimen extraction may alter material properties and conducting robust experiments can be arduous with miniaturised specimens. Additionally, the specimen size effect can restrict spatial resolution of MTT. Indentation hardness testing has greater spatial resolution and is widely used, with many papers published annually. However, substantial finite element analysis is typically required to develop correlations for a given constitutive law, and establishing a widely applicable governing correlation is yet to be done. 
Inverse methods using full-field measurements show promise, particularly the VFM which is less restrictive than the USM and less computationally expensive than FEMU. High spatial resolution characterisation of the unmodified weld specimen is possible using a single test. However, at the time of writing, publications have focused on the characterisation of simple butt-welded joints under uniaxial tensile loading. When the VFM is used, the spatial discretisation of material properties has been defined \textit{a priori} - either as piecewise zones based on microstructure or hardness maps, or as slices assuming homogeneous behaviour through specimen width. Authors have noted that spatial parameterisation of heterogeneous materials is the key issue limiting the characterisation of more complex geometries and loading states. Automated parameterisation would be a breakthrough enabling the characterisation of both more complex geometries, such as welded T-joints, and specimens under more complex loading states, such as bending. 
\section{Extending the virtual fields method for automated spatial parameterisation}
\begin{figure}[h]
	\centering
	\includegraphics[width=\textwidth]{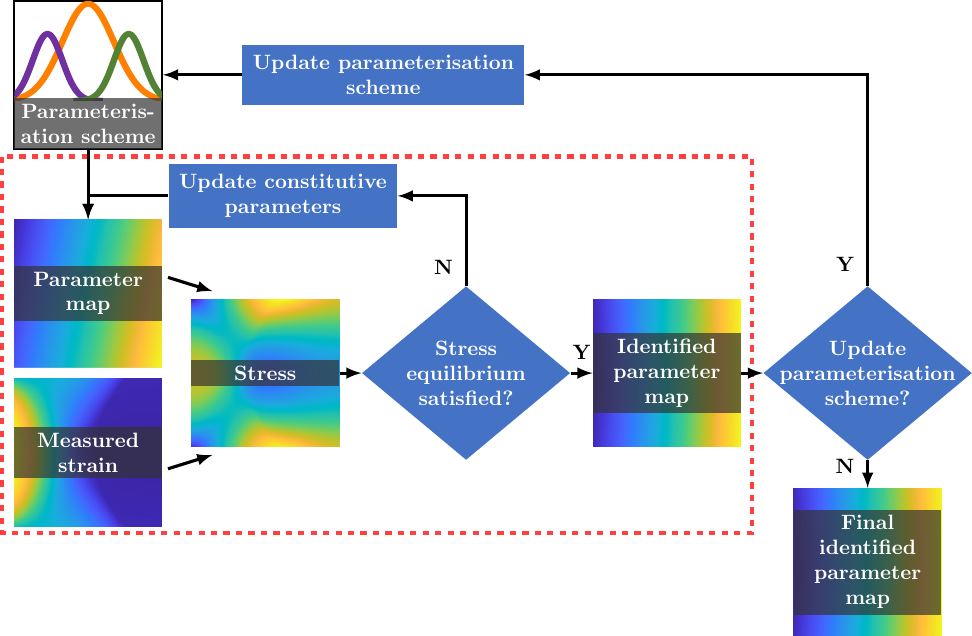}
	\caption[Schematic of the virtual fields method with automated parameterisation.]{Schematic of the virtual fields method with automated parameterisation. The dashed box encloses the core VFM which, as previously described, identifies a map of constitutive parameters that best satisfies stress equilibrium for a particular spatial parameterisation. The outer loop introduces automated parameterisation.} 
	\label{fig:vfm_with_automated_parameterisation}
\end{figure}

For a given set of strain fields and constitutive law, the VFM seeks to identify the constitutive parameters that best satisfy the stress equilibrium of an object. In this work, the strain fields are measured experimentally or assumed to be known. Stress fields can then be reconstructed using a postulated constitutive model and an initial guess of the constitutive parameters. The principle of virtual work (PVW) is then employed to assess the static admissibility of the stress fields, enabling the iteration of constitutive parameters to identify those that best satisfy stress equilibrium. This process is depicted in the dashed red box in \cref{fig:vfm_with_automated_parameterisation}.

As discussed in \cref{sec:intro}, authors have previously used the VFM to identify the heterogeneous mechanical properties of welded joints, however, \textit{a priori} parameterisation was required to discretise the various weld regions. In this work, the VFM is extended to include the automated spatial parameterisation of constitutive parameters as illustrated in \cref{fig:vfm_with_automated_parameterisation}. \cref{sec:parameterisation_of_const_param,sec:stress_equil_metrics} will discuss the spatial parameterisation of the constitutive parameters and the metrics used to assess stress equilibrium, before the assembled identification procedure is outlined in \cref{sec:optimisation_strategy}.

\section{Spatial parameterisation of the constitutive parameters}\label{sec:parameterisation_of_const_param}
In order to reconstruct the stress fields from the measured strain fields, a value for each constitutive parameter must be assigned to every data point, as shown in \cref{eq:stress_parameterisation}. 
\begin{equation}
	{\stress}_{ij}^p = f\left({\strain}_{ij}^{p}, \constitParam_1^{p}, \dots, \constitParam_{\numParam}^{p}\right),
	\label{eq:stress_parameterisation}
\end{equation}
\noindent where the superscript $p$ denotes the $p$\textsuperscript{th} data point; ${\stress}_{ij}^{p}$ and ${\strain}_{ij}^{p}$ are the stress and strain, respectively, and $\constitParam_1^{p}, \dots, \constitParam_{\numParam}^{p}$ are the $\numParam$ constitutive parameters.
Hence, the number of unknowns to be identified is equal to the number of constitutive parameters multiplied by the number of data points. For example, in some experimental work conducted by the author, digital image correlation generated strain fields containing roughly 250,000 data points per image. Assuming a linear hardening constitutive law with only two parameters, yield strength and hardening modulus, the inverse identification problem contains around 500,000 unknowns to be identified. A parameterisation scheme can be used to assign each data point a set of constitutive parameters in a computationally efficient manner, thus reducing the number of unknowns to be identified. Parameterisation strategies are grouped into three categories below: homogeneous regions, mesh-based parameterisation and meshless parameterisation. 

\subsection{Homogeneous region parameterisation}
A simple parameterisation strategy is to discretise the object into uniform regions within which each data point is assigned the same value. For example, in \citep{sutton2008ide}, seven regions were manually defined according to microstructural imaging and previous experimental insight. Each region was then assumed to have homogeneous mechanical properties for identification. Alternative \textit{a priori} insight, such as indentation hardness maps, could be used to the same purpose. However, discretisation through unstructured homogeneous regions is not easily automated for use in heterogeneous parameter identification. 

\subsection{Mesh-based parameterisation}\label{sec:mesh_based_param}
A common discretisation approach in engineering problems is to employ a mesh of elements, for which the values assigned to points within an element are interpolated from nodal values. Hence, the number of unknowns for each constitutive parameter is reduced from the number of data points to the number of nodes. In this work, the mesh elements are typically much larger than the spacing between data points, so a considerable reduction in the number unknowns is obtained. Interpolation from nodal values to data point values is performed using shape functions, with the function's order governing the degree of flexibility in the spatial distribution. Higher order functions provide greater flexibility at the expense of additional degrees of freedom, increasing computational cost and sensitivity to noise. The nodes of adjacent elements can be assigned shared values to enforce continuity, or independent values to permit step changes between elements. Furthermore, as the mesh is solely used to assign values, it can extend beyond the edges of the specimen surface without issue. 

Mesh refinement can be implemented by adjusting the number of elements in the mesh (h-refinement) or changing the order of the shape functions (p-refinement). For example, elements with negligible property variation between them can be merged, and higher-order shape functions can be used in regions of steep property variation. The stress equilibrium metrics discussed in \cref{sec:stress_equil_metrics} could also inform refinement decisions.

A numerical study was conducted to investigate the use of various parameterisation meshes. No refinement was performed, but the meshes for each identification had differing element counts and shape function orders. As expected, the results demonstrated that the rigid structure imposed by mesh-based parameterisation necessitates a rapid increase in the number of unknowns with either p- or h-refinement. However, depending on the distribution of the constitutive parameters to be identified, this increase of unknowns may hinder rather than benefit identification. If not required, these additional unknowns increase optimisation time and can lead to local minima. As the constitutive parameter distribution is unknown in this work, it can be inefficient to enforce the rigid structure imposed using mesh-based parameterisation. As such, the decision was made to consider meshless parameterisation before proceeding with the automated refinement process. 

\subsection{Meshless parameterisation}
Meshless techniques do not rely on any intermediate structure in the spatial domain, enabling efficient utilisation of degrees of freedom in key regions, by providing native sparsity. Two relevant techniques are spline interpolation and basis function interpolation. 

Splines are a set of piecewise polynomials defined using control points to reconstruct smooth functions with minimal curvature. A large variety of spline types exist, with some ensuring interpolated values pass through the control points while others focus on offering more local control of particular regions of the spline. Although future work will investigate the use of splines and other parameterisation strategies, the decision was made to focus here on the use of basis function interpolation.  

Basis function interpolation constructs an interpolant using a weighted sum of basis functions, as in \cref{eq:RBF}.
\begin{equation}
	\interp(x)=\sum_{j=1}^{\numBasis} \rbfWeight_j \kernel(x-\rbfCentre_j),
	\label{eq:RBF}
\end{equation}
\noindent where $\kernel(\cdot)$ is a basis function, $x$ are the evaluation points in the domain, $\rbfCentre_j$ are the \(\numBasis\) centres of the basis functions and $\rbfWeight_j$ are the \(\numBasis\) weights. The interpolant $\interp(x)$ provides the reconstructed field value at any location $x$. In this work, the weights, $\rbfWeight_j$, correspond to the heights of the basis functions and are treated as unknowns to be optimised during the identification process.

\begin{figure}[h!]
	\centering
	\begin{subfigure}{0.75\textwidth}
		\includegraphics[width=\textwidth]{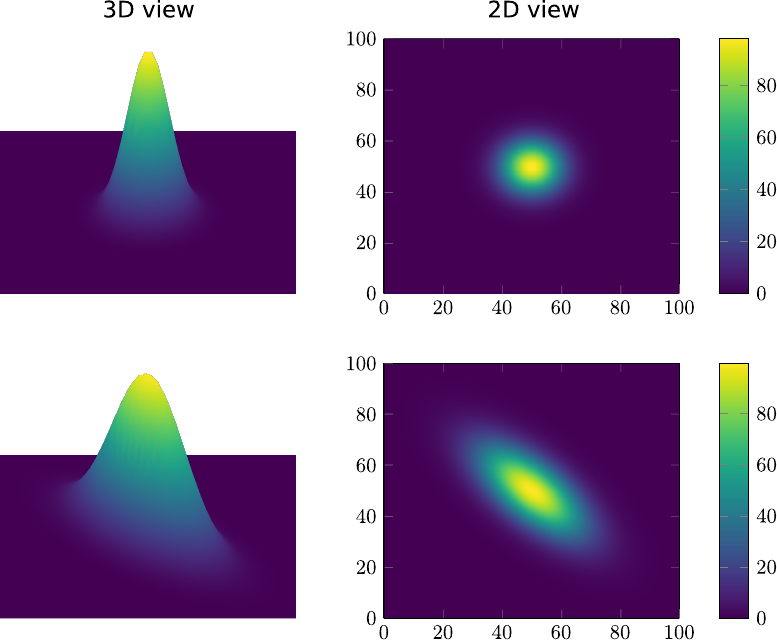}
	\end{subfigure} \\
	\begin{subfigure}{0.75\textwidth}
		\vspace{8mm} 
		\centering
		\small 
		\begin{tblr}{
				colspec = { 
					X[1.9,c,b]  
					X[c,b]      
					X[c,b]      
					X[c,b]      
					X[1.1,c,b]  
					X[1.1,c,b]  
					X[c,b]      
				},
				width = \textwidth,
				row{1} = {font=\bfseries\boldmath},  
				hline{1,4} = {1.2pt},  
				hline{2} = {0.5pt},    
			}
			 & X-Coord & Y-Coord & Height & Variance in x & Variance in y & Angle \\
			upper figures & 50 & 50 & 100 & 100 & 100 & 0 \\
			lower figures & 50 & 50 & 100 & 100 & 500 & 30 \\
		\end{tblr}
		\label{tab:gaussian_rbf_example_parameters}
	\end{subfigure}
	\captionsetup{width=0.75\textwidth}
	\caption[Visualisation of univariate and bivariate Gaussian basis functions.]{Visualisation of the univariate (upper) and bivariate (lower) Gaussian basis functions defined using the table parameters.}
	\label{fig:rbf_example_figures}
\end{figure}

Many forms of basis functions exist, but Gaussian basis functions were selected for this work due to their popularity and ability to efficiently reconstruct surfaces with a large range of spatial frequencies. \cref{eq:RBF_Gaussian} defines the Gaussian basis function, which has a maximum of four or six unknowns, for a univariate or bivariate function respectively, as illustrated in \cref{fig:rbf_example_figures}. 

\begin{equation}
	\kernel(x) = \exp\left( -\frac{1}{2} (x - \rbfCentre_j) \covMat^{-1} (x - \rbfCentre_j)^\top \right),
	\label{eq:RBF_Gaussian}
\end{equation}
\noindent where $\kernel$ is the Gaussian basis function centred at point $\rbfCentre_j$, and $\covMat$ is the covariance matrix, defined in \cref{eq:RBF_Gaussian_covariant}.

\begin{equation}
	\covMat = \begin{bmatrix}
		\cos(\rotAngle) & -\sin(\rotAngle) \\
		\sin(\rotAngle) & \cos(\rotAngle)
	\end{bmatrix} 
	\begin{bmatrix}
		\rbfVarOne & 0 \\
		0 & \rbfVarTwo
	\end{bmatrix}
	\begin{bmatrix}
		\cos(\rotAngle) & -\sin(\rotAngle) \\
		\sin(\rotAngle) & \cos(\rotAngle)
	\end{bmatrix}^\top ,
	\label{eq:RBF_Gaussian_covariant}
\end{equation}
\noindent where $\rbfVarOne$ and $\rbfVarTwo$ are the variances along the principal axes and $\rotAngle$ is the rotation angle of the principal axes relative to the coordinate system.

\begin{figure}[h]
	\centering
	\includegraphics[width=10cm]{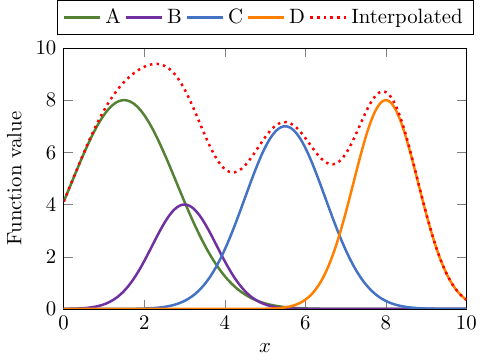}
	\caption{Plot showing the superposition of 1D Gaussian basis functions.}
	\label{fig:RBF_Superposition}
\end{figure}

The concept is most easily demonstrated in one dimension. \cref{fig:RBF_Superposition} visualises the superposition of four Gaussian basis functions to generate an interpolation function. After selecting the basis function type, the quantity, shape and position of the basis functions are optimised to reconstruct the parameter maps that result in the minimum error in stress equilibrium. As such, suitable metrics are required to assess stress equilibrium. These metrics are discussed in the next section.

\section{Metrics to assess stress equilibrium}\label{sec:stress_equil_metrics}
In this work, stress equilibrium is evaluated using three metrics, each derived from a different formulation of virtual fields within the principle of virtual work. As such, a brief outline of the relevant VFM theory is presented below before the metrics are introduced.
	
\subsection{The principle of virtual work (PVW)}{\label{sec:The Principle of Virtual Work}}
\cref{eq:stress_equilibrium} shows the local stress equilibrium equation, which must be satisfied at every point in a continuum.
\begin{equation}
\stress_{ij,j} + \bodyforce_i = \density \accel_i.
\label{eq:stress_equilibrium}
\end{equation}
Einstein notation is used herein, for which repeated indices indicate summation and a comma indicates the spatial derivative.  $\stress_{ij,j}$ is the divergence of the Cauchy stress tensor, $\bodyforce_i$ is the body force vector, $\density$ is the density and $\accel_i$ is the acceleration vector. Multiplying both sides of the stress equilibrium equation with a continuous and piecewise-differentiable vectorial test function, $\vdisp_i^*$, and integrating over the volume under investigation, $V$, results in \cref{eq:equil_eq_vectorial_func}.
\begin{equation}
\int_V \stress_{ij,j}\,\vdisp_i^*\,\mathrm{d}V
+\int_V \bodyforce_i\,\vdisp_i^*\,\mathrm{d}V
=
\int_V \density \accel_i\,\vdisp_i^*\,\mathrm{d}V.
\label{eq:equil_eq_vectorial_func}
\end{equation}
Cauchy's lemma, $\traction_i=\stress_{ij} \normal_j$, relates the stress tensor and the traction vector $\traction_i$ at the boundary surface, where $\normal_j$ is the outward facing unit vector  of the surface. Using integration by parts, the divergence theorem, and Cauchy's lemma, the first term of \cref{eq:equil_eq_vectorial_func} can be separated to produce \cref{eq:PVW}. 
\begin{equation}
- \int_V \stress_{ij}\,\vdisp_{i,j}^*\,\mathrm{d}V
+\int_{S_f} \traction_i\,\vdisp_i^*\,\mathrm{d}S
+\int_V \bodyforce_i\,\vdisp_i^*\,\mathrm{d}V
=
\int_V \density \accel_i\,\vdisp_i^*\,\mathrm{d}V.
\label{eq:PVW}
\end{equation}
\noindent where $S_f$ is the portion of the boundary surface, $S$, over which the traction vector, $\traction_i$ acts. \cref{eq:PVW} is the foundational equation for the PVW. It expresses the stress equilibrium equation in the weak form (accounting for the global equilibrium over the volume). In other words, the PVW is mathematically equivalent to the local equilibrium equation (with the force boundary conditions accounted for).  It is also worth noting that this equation is independent of the selected constitutive model.

The first term in \cref{eq:PVW} is referred to as the virtual work done by internal forces or the internal virtual work (IVW). The second and third terms are referred to as the virtual work done by external forces or the external virtual work (EVW). The right-hand side term is the virtual work done by the acceleration, or the inertial virtual work. 

The term $\vdisp_i^*$ is a test function known as the virtual displacement. This term does not require physical interpretation and acts solely as a spatial weighting function over the body. The associated virtual strain tensor is defined as $\strain^*_{ij} = \frac{1}{2}\left(\vdisp^*_{i,j} + \vdisp^*_{j,i}\right)$. The virtual strains are unrelated to the physical strains and only act as spatial weighting functions. Under quasi-static conditions, statically admissible stress fields satisfy the PVW for any choice of continuous and piecewise-differentiable virtual field. These virtual field terms are discussed further below. 

\paragraph{PVW for surface measurements, quasi-static loading and negligible body forces}
Assuming quasi-static loading and negligible body forces, the terms containing $\accel_i$ and $\bodyforce_i$ can be neglected. Furthermore, optical methods such as digital image correlation (DIC) only provide 2D surface measurements. Therefore, the VFM requires assumptions about the through-thickness material behaviour. In this work, thin specimens are used and the stress is assumed constant throughout the thickness, $\thickness$. Note, it is possible to correct for small amounts of out-of-plane bending (see Appendix C in \citep{peshave2026dic}).
\begin{equation}
-\,\thickness\int_{S}\stress_{ij}\,\strain_{ij}^*\,\mathrm{d}S
+ \thickness\int_{L_f} \traction_i\,\vdisp_i^*\,\mathrm{d}L
=0.
\label{eq:PVW_2DQS}
\end{equation}

\paragraph{Simplifying unknown traction distributions}{\label{sec:simplifying traction distributions}}
The second term (EVW) in \cref{eq:PVW_2DQS} requires knowledge of the tractions, $\traction_i$, acting on the specimen boundary. However, during experimentation the resultant forces are usually measured, whereas the exact traction distribution is generally unknown. Defining a constant virtual displacement, $\vdisp_i^*$, over the boundary surface with the unknown traction distribution, $S_f$, allows the constant term $\vdisp_i^*$ to be factored outside the integral. This enables the EVW term to be calculated from the measured resultant applied force, $\resForce_i$, as in \cref{eq:ExternalWorkAppliedForce}.
\begin{equation}
\int_{S_f} \traction_i\,\vdisp_i^*\,\mathrm{d}S
=
\vdisp_i^*\int_{S_f} \traction_i\,\mathrm{d}S
=
\vdisp_i^* \resForce_i,
\label{eq:ExternalWorkAppliedForce}
\end{equation}

\paragraph{PVW using discrete measurements}
The strain data obtained using DIC will take the form of discrete data points each having a strain value, $\strain_{ij}^p$, and a surface area, $\areaPoint^p$. The full-field data should have sufficient spatial resolution to approximate the surface integral as a discrete sum using the mid-point rule (the validity of this approximation can be assessed using synthetic image deformation). Hence, the PVW can be rewritten as in \cref{eq:PVW_2DQS_discrete}.
\begin{equation}
\underbrace{
-\,\thickness \sum_{p=1}^{\numPoints}
\stress_{ij}^{p}\,\strain_{ij}^{*p}\,\areaPoint^p
}_{\mathrm{IVW}}
+
\underbrace{
\vdisp_i^* \resForce_i
\vphantom{\sum_{p=1}^{\numPoints}
\stress_{ij}^{p}\,\strain_{ij}^{*p}\,\areaPoint^p}
}_{\mathrm{EVW}}
= 0 .
\label{eq:PVW_2DQS_discrete}
\end{equation}
\noindent where the superscript $p$ denotes the $p$\textsuperscript{th} data point; $\numPoints$ is the number of discrete data points; $\thickness$ is the specimen thickness; $\stress_{ij}^p$ and $\strain_{ij}^{*p}$ are the stress and virtual strain, respectively; $\areaPoint^p$ is the surface area associated with point $p$; $\vdisp_i^*$ is the virtual displacement on the traction boundary; and $\resForce_i$ is the resultant applied force on the traction boundary.

The virtual fields act as spatial filters -- weighting the cost function. For statically admissible stress fields, the PVW is satisfied for any choice of continuous and piecewise-differentiable virtual field. Hence, the choice of virtual fields is highly flexible and for homogeneous identification in the absence of noise and with a correctly postulated constitutive equation, any set of virtual fields should result in the same identified parameters. In reality, noise and model error exist. By selecting virtual fields that assign more weight to regions with higher signal-to-noise ratio, the accuracy of the identification can be enhanced. 

\subsection{Sensitivity based virtual fields}
\cref{eq:PVW_2DQS_discrete} can be reformulated into a cost function suitable for elastoplastic parameter identification, as shown in \cref{eq:sbvf_cost_function} \citep{marek2017sen}. This cost function evaluates the squared residual, $\sbvfCost$, between the internal and external virtual work over all time steps, $t = 1, \dots, \numSteps$, and all selected virtual fields, $f = 1, \dots, \numVF$.
\begin{equation}
\sbvfCost
=
\sum_{f=1}^{\numVF}
\sum_{t=1}^{\numSteps}
\left[
\thickness \sum_{p=1}^{\numPoints}
\stress_{ij}^{p}(t)\,\strain_{ij}^{*p,f}(t)\,\areaPoint^p
-
\vdisp_i^{*,f}(t)\,\resForce_i(t)
\right]^2 .
\label{eq:sbvf_cost_function}
\end{equation}

\citeauthor{marek2017sen} developed the sensitivity-based virtual fields (SBVFs), to automate the generation of virtual fields preferentially weighting regions (in space and time) that contain information about the parameter being identified \citep{marek2017sen}. To do this, stress sensitivity maps are computed for each degree of freedom being identified. The stress sensitivities are established by perturbing the value of the current unknown and calculating the resulting change in stress. The SBVF's are the virtual fields that replicate the stress sensitivity maps as closely as possible, while also enforcing any required virtual boundary conditions. 

\subsection{Force reconstruction error}\label{sec:RAFE}
\begin{figure}[h!]
	\centering
	\includegraphics[width=0.25\textwidth]{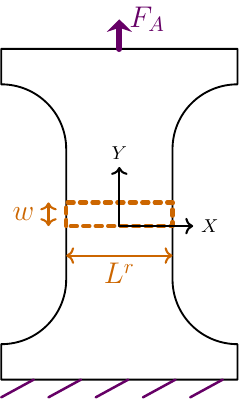}
	\caption{Schematic of slice used for the force reconstruction error virtual fields.}
	\label{fig:sliceVF_schematic}
\end{figure}

\cref{fig:sliceVF_schematic} shows a schematic of a typical tensile test. The applied longitudinal force, $\appliedForce$, is measured and must be in equilibrium with the average longitudinal stress at each cross-section of the specimen. The piecewise virtual fields shown in \cref{eq:slice_virtDisp,eq:slice_virtStrain} are defined to negate the contribution to internal virtual work of all data outside the dashed region (slice), and the external virtual work only involves the applied force. These virtual fields have been detailed previously in \citep{sutton2008ide,lelouedec2013ide, peshave2024met}, but are outlined below for convenience. 

\begin{minipage}{0.45\textwidth}
	\begin{equation}
		\vdisp^*_{x}=
		\begin{cases}
			0 & \text{if } y<0   \nonumber \\
			0 & \text{if } 0<y<w \nonumber \\
			0 & \text{if } y>w \nonumber \\
		\end{cases}
	\end{equation}
\end{minipage}
\begin{minipage}{0.45\textwidth}
	\begin{equation}
		\vdisp^*_{y}=
		\begin{cases}
			0 & \text{if } y<0    \\
			y & \text{if } 0<y<w  \\
			w & \text{if } y>w   \\
		\end{cases}
		\label{eq:slice_virtDisp}
	\end{equation}
\end{minipage}

\begin{minipage}{0.33\textwidth}
	\begin{equation}
		\strain_{xx}^*=
		\begin{cases}
			0 & \text{if } y<0   \nonumber \\
			0 & \text{if } 0<y<w \nonumber \\
			0 & \text{if } y>w \nonumber \\
		\end{cases}
	\end{equation}
\end{minipage}
\begin{minipage}{0.33\textwidth}
	\begin{equation}
		\strain_{yy}^*=
		\begin{cases}
			0 & \text{if } y<0   \nonumber \\
			1 & \text{if } 0<y<w \nonumber \\
			0 & \text{if } y>w \nonumber \\
		\end{cases}
	\end{equation}
\end{minipage}
\begin{minipage}{0.33\textwidth}
	\begin{equation}
		\strain_{xy}^*=
		\begin{cases}
			0 & \text{if } y<0    \\
			0 & \text{if } 0<y<w  \\
			0 & \text{if } y>w  \\
		\end{cases}
		\label{eq:slice_virtStrain}
	\end{equation}
\end{minipage}
\vspace{1em}

Substituting these virtual fields into the PVW (\cref{eq:PVW_2DQS_discrete}) results in the following virtual work terms. 
\begin{equation}
	\begin{split}
		\mathrm{IVW}^{r}
		&=
		-\,\thickness\int_{S^r}\stress_{ij}\,\strain_{ij}^*\,\mathrm{d}S \\
		& \approx\;
		-\,\thickness\,w\,L^{r}\,\overline{\stress}_{yy}^{\,r} \\
		& \approx\;
		-\,w\,\reconForce^{\,r},  \\
	\end{split}
\label{eq:slice_vf_ivw}
\end{equation}
\noindent where $r = 1, \dots, \numSlices$ denotes the slice index, $\numSlices$ is the total number of slices, and $S^r$ denotes the $r$\textsuperscript{th} slice region (with area $wL^r$). $w$ is the slice width, $L^r$ the slice length and $\overline{\stress}_{yy}^r$ is the mean longitudinal stress of the data points within slice $r$. $\reconForce^r$ denotes the reconstructed force over the slice. In the examples below, each slice was defined to have a width of 5 data points ($w=0.8~mm$) to provide some smoothing whilst still maintaining good spatial resolution. 
\begin{equation}
	\begin{split}
		\mathrm{EVW}^{r}
		&=
		\vdisp_i^*\resForce_i \\
		&\approx
		w\,\appliedForce,
	\end{split}
\label{eq:slice_vf_evw}
\end{equation}
\noindent where $\appliedForce$ is the resultant force applied at the traction boundary. 

Recombining the internal and external virtual work terms therefore enforces that the reconstructed force, $\reconForce^r$, equals the applied force, $\appliedForce$, when the slice region is in equilibrium. Deviation from this condition provides a measure of equilibrium error for each slice region. This error can be expressed with units of force or as a dimensionless term, which is useful for assessing the relative equilibrium error of slices at different load steps. In this work, the Force Reconstruction Error (FRE) is defined as a dimensionless term as shown in \cref{eq:RAFR}.
\begin{equation}
	\begin{split}
		\mathrm{FRE}^{r}
		&=
		\frac{\reconForce^{\,r}}{\appliedForce}-1\\
		&=
		\frac{\thickness\,L^{r}\,\overline{\stress}_{yy}^{\,r}}{\appliedForce}-1 .\\
	\end{split}
\label{eq:RAFR}
\end{equation}
It is possible to formulate the FRE metric as a cost function to be minimised, for which the corresponding constitutive parameters for each slice are identified independently (as in \citep{lelouedec2013ide}). Alternatively, the spatial-temporal fields can be aggregated using statistics such as the root mean square (RMS). In this work, a force-weighted RMS is used to reduce the influence of low-force load steps, which have a lower signal-to-noise ratio. Analysis using known, numerically defined stress data indicated that the variance of the relative force reconstruction error was inversely proportional to the square of the applied force. This same scaling was observed for the equilibrium gap indicator term discussed later. Hence, temporal weights proportional to the square of the applied force are used, defined as 
\begin{equation}
\wgt(t)=
\frac{\appliedForce(t)^2}{\frac{1}{\numSteps}\sum_{t=1}^{\numSteps}\appliedForce(t)^2}.
\label{eq:force_weight}
\end{equation}
where $\numSteps$ is the number of load steps and $t = 1, \dots, \numSteps$ denotes the step index. The weights are normalised to have a mean value of one so that the weighted RMS remains comparable in magnitude to an unweighted RMS.

The weighted temporal RMS (\cref{eq:FRE_weighted_temporal_rms}) is computed for each region $r$ to provide information on how the FRE varies spatially. The force-weighted spatial-temporal RMS (\cref{eq:FRE_weighted_spatiotemporal_rms}) returns a single scalar aggregated over all slices, for use in the optimisation process.
\begin{equation}
\mathrm{FRE}_{\mathrm{weighted\_temporal\_RMS}}^{\,r}
=
\sqrt{
\frac{1}{\numSteps}
\sum_{t=1}^{\numSteps}
\wgt(t)\,
\left[\mathrm{FRE}^{\,r}(t)\right]^2
}.
\label{eq:FRE_weighted_temporal_rms}
\end{equation}
\begin{equation}
\mathrm{FRE}_{\mathrm{weighted\_spatial\text{-}temporal\_RMS}}
=
\sqrt{
\frac{1}{\numSlices}
\sum_{r=1}^{\numSlices}
\left(
\frac{1}{\numSteps}
\sum_{t=1}^{\numSteps}
\wgt(t)\,
\left[\mathrm{FRE}^{\,r}(t)\right]^2
\right)
}.
\label{eq:FRE_weighted_spatiotemporal_rms}
\end{equation}

\subsection{Equilibrium gap indicator}
Another metric -- the Equilibrium Gap Indicator (EGI) -- examines local discrepancies in stress equilibrium within an isolated window of data points. \citeauthor{devivier2013imp} used the EGI to detect damage in composite materials, and \citeauthor{considine2017smo} subsequently used the EGI formulation to assess local equilibrium in specimens of smoothly varying stiffness \citep{devivier2013imp,considine2017smo}. More recently, \citeauthor{peshave2024met} used this metric in combination with the FRE to evaluate constitutive model fitness \citep{peshave2024met}. 

\begin{figure}[h!]
    \centering
	\hspace*{0\textwidth}%
    \subfloat[\centering mesh]{%
        \includegraphics[width=0.26 \textwidth]{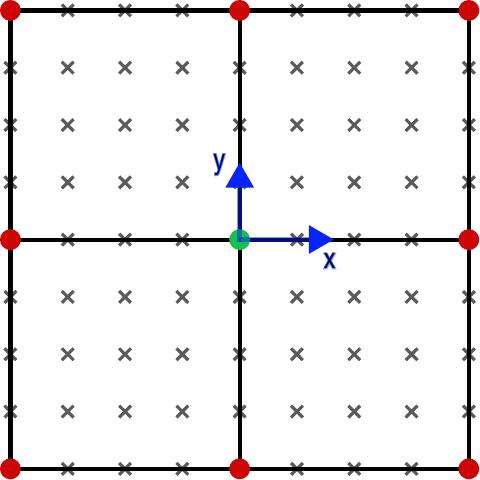}}%
	\hspace*{0.08\textwidth}%
    \subfloat[\centering $\vdisp^*_{x}$]{%
        \includegraphics[width=0.32 \textwidth]{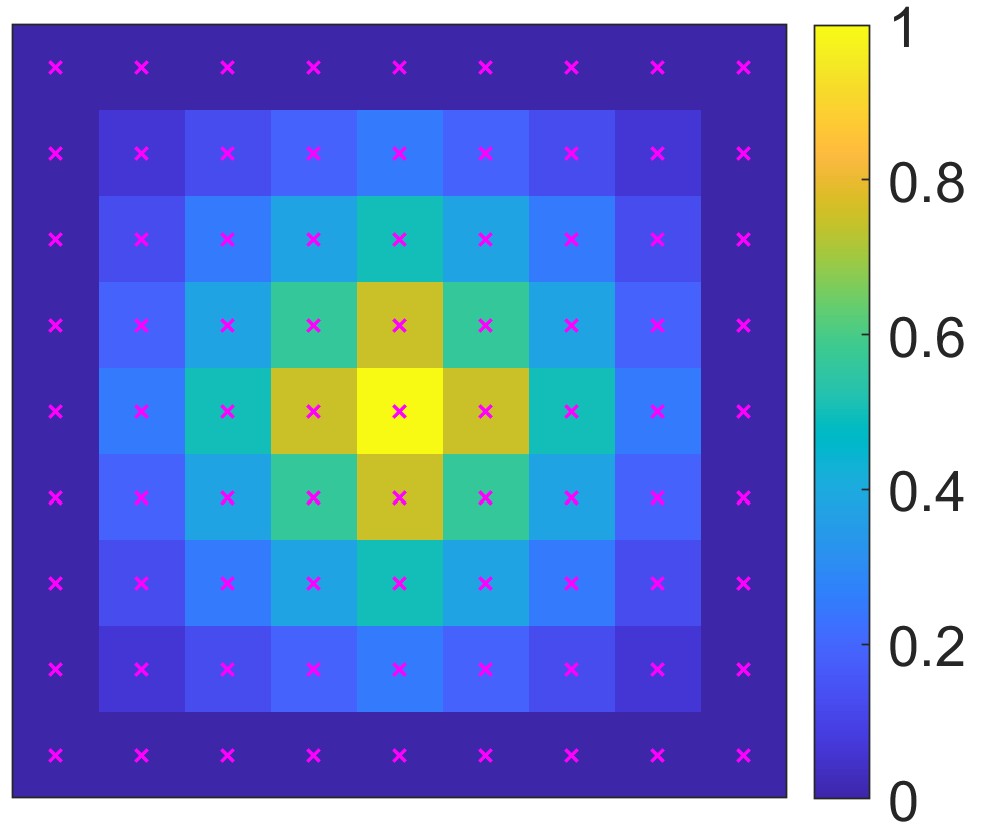}}%
	\hspace*{0.02\textwidth}%
    \subfloat[\centering $\vdisp^*_{y}$]{%
        \includegraphics[width=0.32 \textwidth]{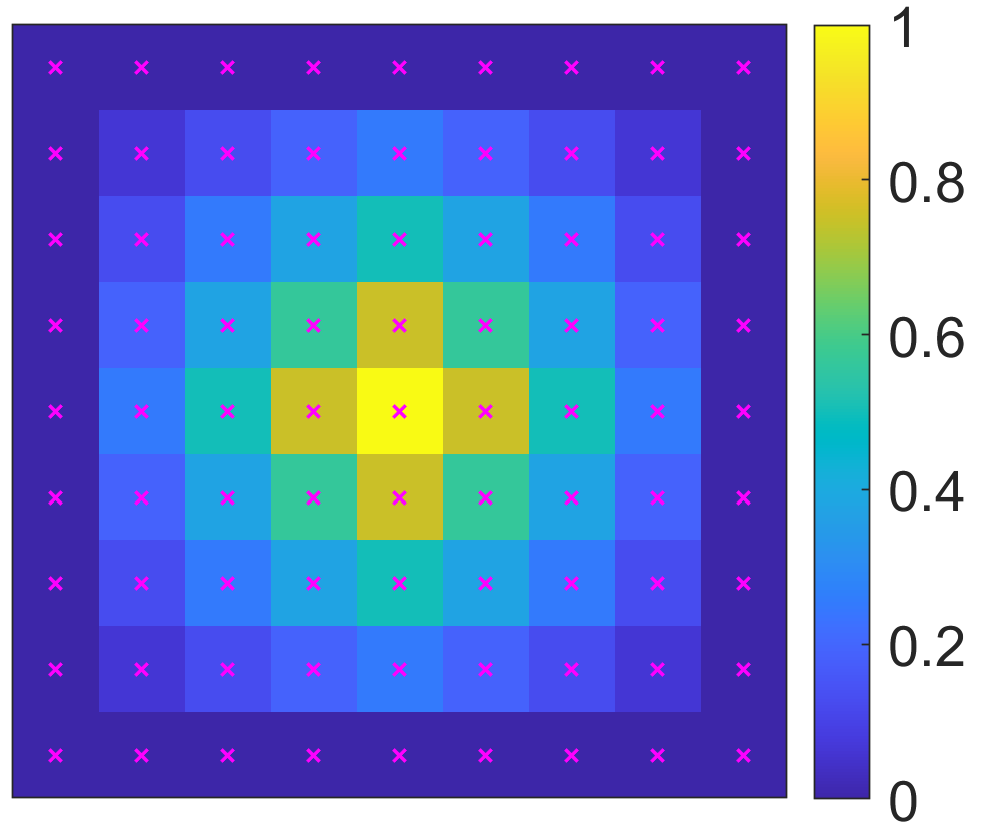}}%
	\hspace*{\fill}%
    \vskip\baselineskip 
    %
    \subfloat[\centering $\strain_{xx}^*$]{%
        \includegraphics[width=0.32\textwidth]{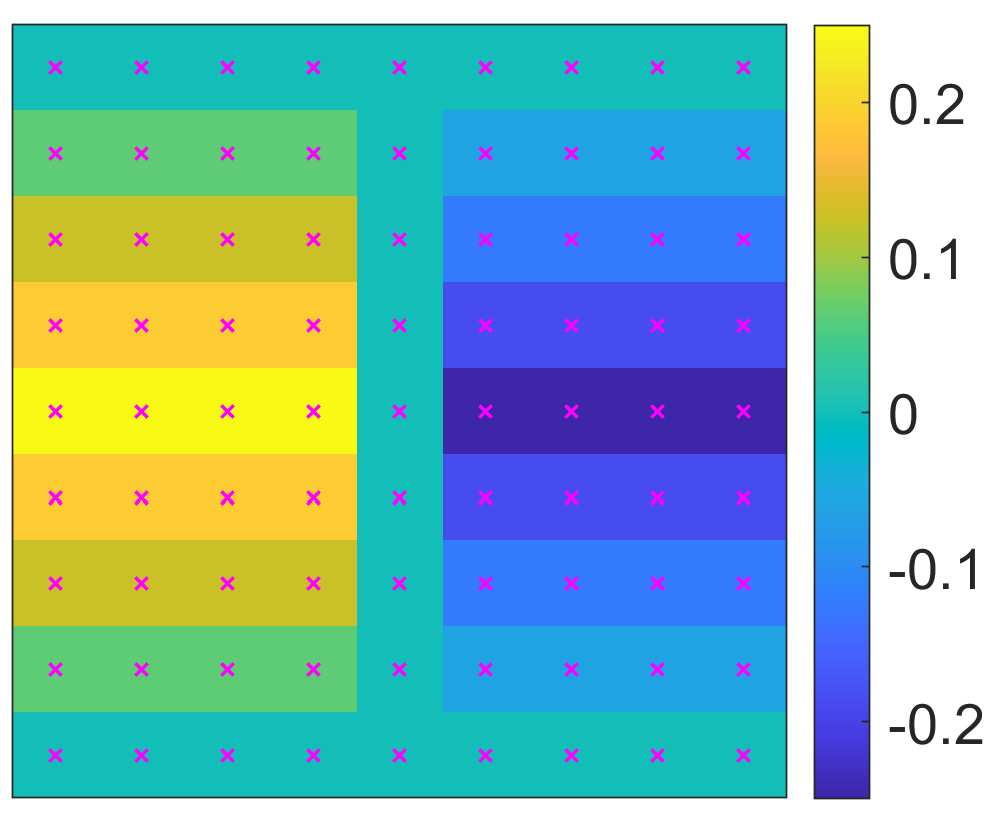}}%
    \hfill
    \subfloat[\centering $\strain_{yy}^*$]{%
        \includegraphics[width=0.32\textwidth]{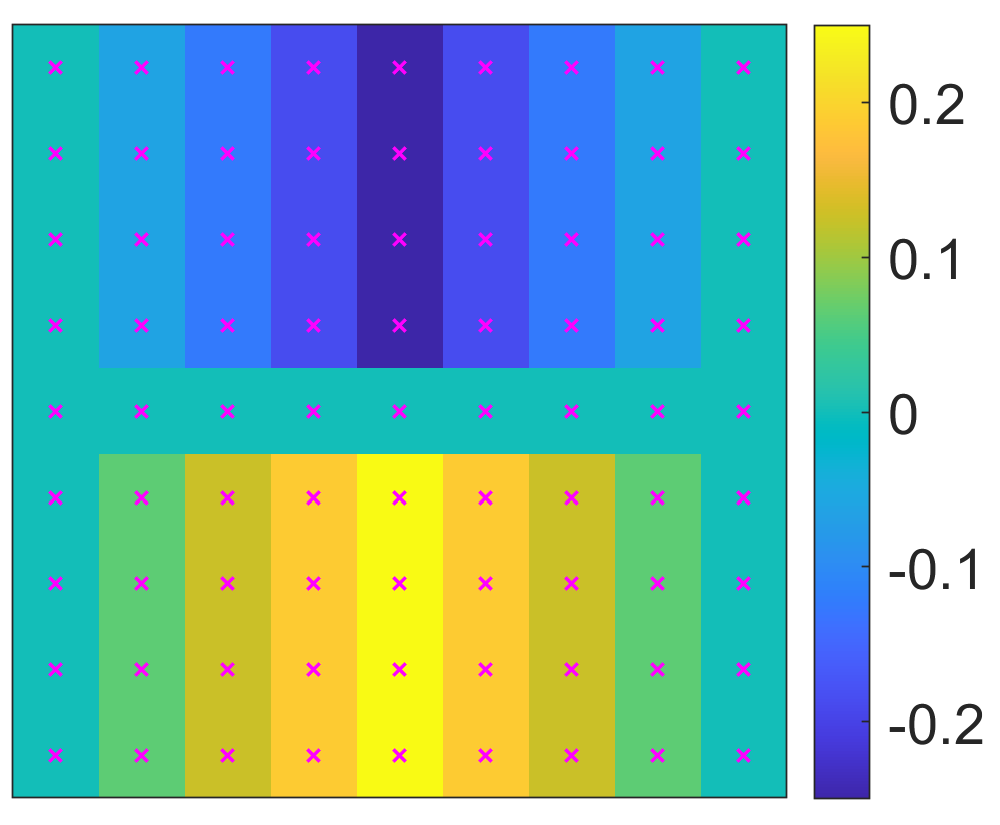}}%
    \hfill
    \subfloat[\centering $\strain_{xy}^*$]{%
        \includegraphics[width=0.32\textwidth]{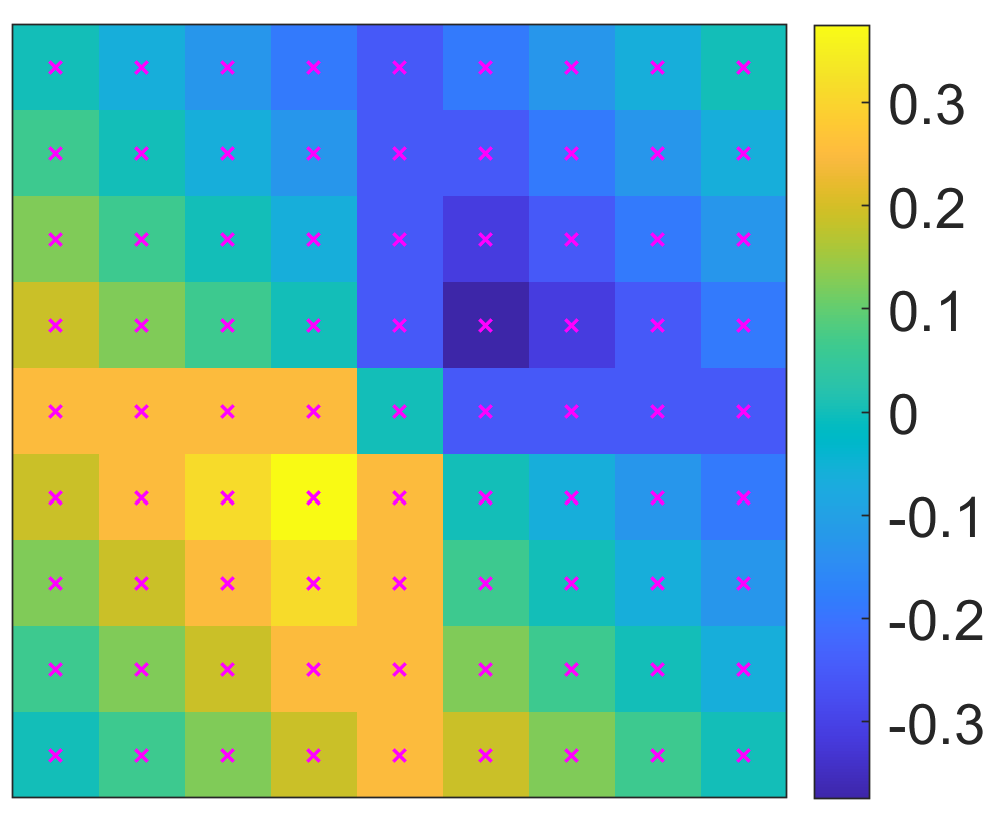}}%
    \caption{Schematic of the virtual mesh used to construct equilibrium gap virtual fields, and associated virtual displacements and strains for the [1,1] virtual displacement of the central node.}
    \label{fig:EG__VF_combined}
\end{figure}

A mesh consisting of four elements and nine nodes is used to define a piecewise virtual field for a given EGI inspection window (see \cref{fig:EG__VF_combined}). The window size is user-defined, and is discussed further below. In order to nullify the contribution of any unknown tractions on the window boundary, the eight boundary nodes are assigned to have a virtual displacement of zero. In this work, the centre node is assigned one of two independent virtual displacement vectors: $[1,1]$ or $[1,-1]$. Any other virtual movement of the central node will be a linear combination of these two vectors and will not add any information to the process. The virtual displacement and strain fields, for a virtual displacement vector of $\vdisp^*=[1,1]$ are shown in \cref{fig:EG__VF_combined}.

As a first attempt, only EGI values computed using the $\vdisp^*=[1,1]$ virtual displacement vector were used. However, for certain geometry and load configurations, an improvement in the metric performance was noted when both sets of virtual displacements ($\vdisp^*=[1,1]$ and $\vdisp^*=[1,-1]$) were used and the EGI values averaged. Future work could explore using virtual meshes with more elements to enable more complex virtual fields.

\begin{figure}[h]
	\centering
	\includegraphics[width=\textwidth]{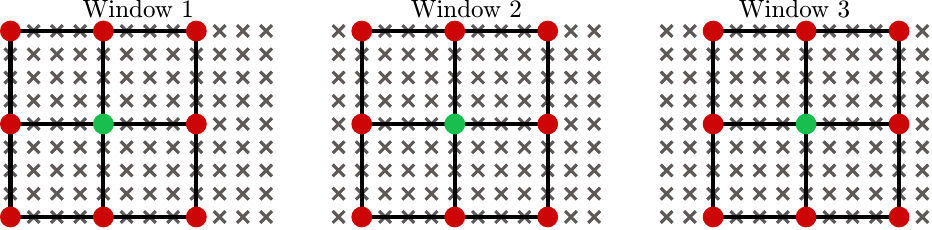}
	\caption{Schematic showing the rastering of the EGI window across the data.}
	\label{fig:EG_window_rastering}
\end{figure}

As the virtual displacements at the traction boundaries are all set to zero, the EVW term disappears from the PVW (\cref{eq:PVW_2DQS_discrete}) resulting in \cref{eq:EG_pvw}.
\begin{equation}
-\,\thickness \sum_{p=1}^{\numPoints^m}
\stress_{ij}^{p}\,\strain_{ij}^{*p}\,\areaPoint^p
=0 .
\label{eq:EG_pvw}
\end{equation}
\noindent where $m = 1, \dots, \numWin$ denotes the window index, $\numWin$ is the total number of windows, and $\numPoints^m$ is the number of data points in the EGI window at window position $m$. $\stress_{ij}^{p}$ and $\strain_{ij}^{*p}$ are the stress and virtual strain, respectively; and $\areaPoint^p$ is the surface area associated with point $p$.
The left-hand side of \cref{eq:EG_pvw} should yield zero when the stress field satisfies equilibrium. Hence, any deviation from equilibrium can be written in terms of the Equilibrium Gap Indicator (EGI), defined as
\begin{equation}
\mathrm{EGI}_{\mathrm{raw}}^{m}
=
\thickness \sum_{p=1}^{\numPoints^m}
\stress_{ij}^{p}\,\strain_{ij}^{*p}\,\areaPoint^p.
\label{eq:EGI_raw}
\end{equation}
This EGI metric is computed for all data points by rastering the virtual window across the data, as depicted in \cref{fig:EG_window_rastering}. Each evaluation returns a scalar value -- equal to zero if stress equilibrium is satisfied. This scalar value is assigned to the centre data point of the window, and the window is then moved by a user-defined step to its next location. Throughout this work, an EGI step size of one data point was used in both directions; this maximises the spatial resolution of the metrics at the expense of computation time. 

To aid comparison of EGI metrics computed using multiple window sizes across multiple load steps, the EGI metric is normalised by the number of points in the EGI window and the applied force, as shown in \cref{eq:EGI_normalised}. Normalisation using the root mean square of the equivalent stress was also trialled, however, this biased identification towards constitutive parameters that resulted in higher stress fields and hence lower EGI values.

\begin{equation}
\mathrm{EGI}^{m}_{\mathrm{normalised}}
=
\frac{\mathrm{EGI}_{\mathrm{raw}}^{m}}{\numPoints^{m}\,\appliedForce} .
\label{eq:EGI_normalised}
\end{equation}
Evaluating $\mathrm{EGI}_{\mathrm{raw}}^{m}$ or $\mathrm{EGI}^{m}_{\mathrm{normalised}}$ for each window position and time step results in a spatial-temporal field of equilibrium gap values. As with the FRE RMS terms in \cref{sec:RAFE}, a force-weighted temporal RMS and a force-weighted spatial-temporal RMS are computed using the weights defined in \cref{eq:force_weight}. A brief investigation explored whether only using data from specific steps, such as steps with plastic deformation or steps with larger error values, improved identification. Results were inconclusive, so in this work all time steps were used for simplicity.

\begin{equation}
\mathrm{EGI}_{\mathrm{weighted\_temporal\_RMS}}^{m}
=
\sqrt{
\frac{1}{\numSteps}
\sum_{t=1}^{\numSteps}
\wgt(t)\,
\left[\mathrm{EGI}^{\,m}_{\mathrm{normalised}}(t)\right]^2
}.
\label{eq:EGI_weighted_temporal_rms}
\end{equation}
\begin{equation}
\mathrm{EGI}_{\mathrm{weighted\_spatial\text{-}temporal\_RMS}}
=
\sqrt{
\frac{1}{\numWin}
\sum_{m=1}^{\numWin}
\left(
\frac{1}{\numSteps}
\sum_{t=1}^{\numSteps}
\wgt(t)\,
\left[\mathrm{EGI}^{m}_{\mathrm{normalised}}(t)\right]^2
\right)
}.
\label{eq:EGI_weighted_spatiotemporal_rms}
\end{equation}

\subsection{Summary of metrics to assess stress equilibrium}
The SBVF's evaluate virtual work terms globally while assigning more weight to regions with higher signal-to-noise for each parameter being identified. However, in specimens with heterogeneous mechanical properties, compensation can occur, whereby a region of higher stress balances a region of lower stress. In such cases, evaluating the virtual work terms globally across the specimen could lead to local minima. The EGI and FRE metrics assess stress equilibrium more locally. The EGI assesses how well the spatial distribution of stress within each window satisfies static admissibility. However, it does not account for the magnitude of the applied force and is therefore insensitive to uniform scaling of the stress field. As such, all data points in the stress field may be incorrect by some factor and the EGI will not detect any issue provided the stress distribution is balanced. On the other hand, the FRE only considers the longitudinal component of stress and provides no discretisation throughout the specimen cross-section. However, it returns a value directly comparable with the magnitude of applied force. 

Each of these metrics are different but complementary. When combined, these metrics provide rich spatial insight into the distribution and magnitude of stress equilibrium throughout the specimen. Using a combination of different virtual fields enhances identification. The next section discusses how these three metrics are utilised for the inverse identification methodology. 
\section{Identification procedure and verification}\label{sec:optimisation_strategy}
This section details the assembly of the parameterisation methods and equilibrium metrics to automate the spatial mapping of heterogeneous elastoplastic properties, as depicted in \cref{fig:vfm_with_automated_parameterisation}. The developed approach is demonstrated on synthetic data, mimicking data obtained experimentally.

\subsection{Numerical toolchain}\label{sec:numerical_toolchain}

\begin{figure}
	\centering
	\includegraphics[width=\linewidth]{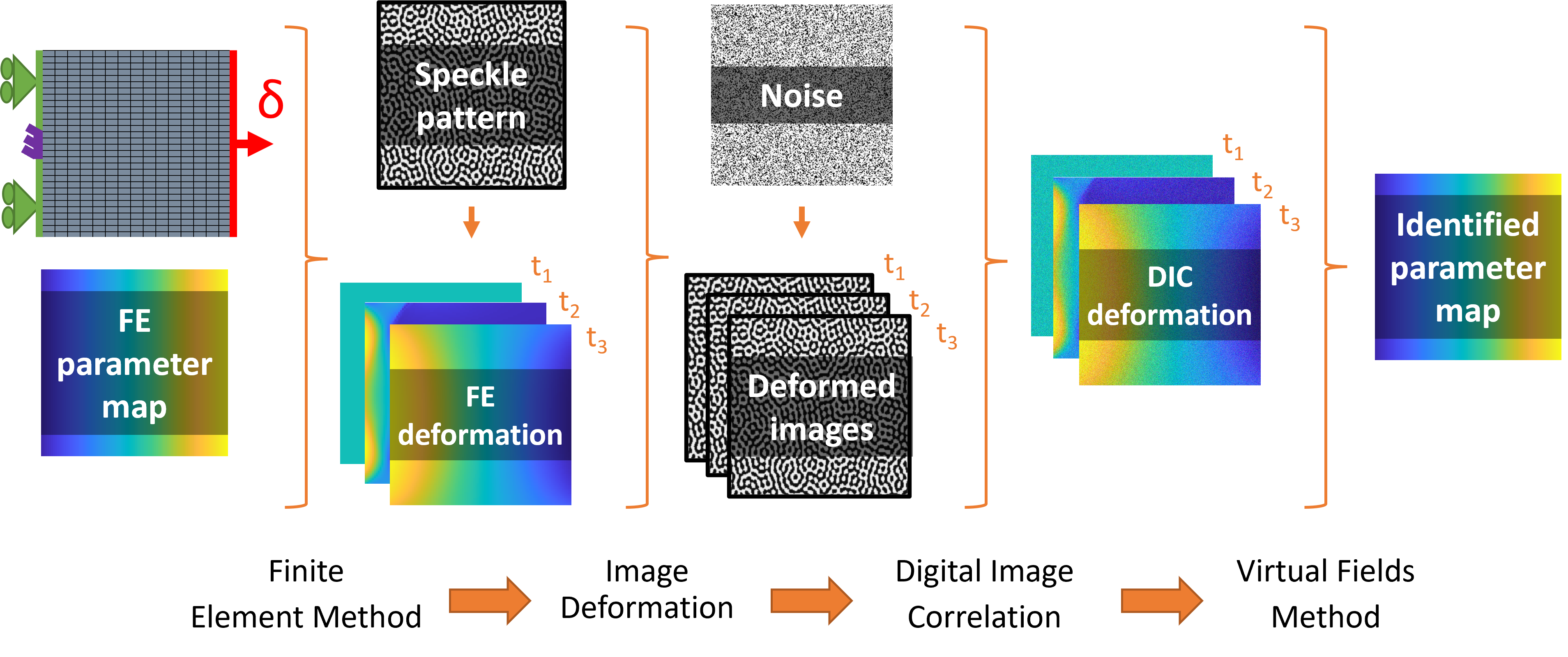}
	\caption{Schematic of numerical toolchain used for the development of the methodology.}
	\label{fig:toolchain}
\end{figure}

To accelerate the development and verification process, a toolchain, shown in \cref{fig:toolchain},  was developed which encapsulates each step of the data generation and identification process. 

Finite element (FE) analysis was used to generate deformation data for a specified geometry and boundary conditions. In this work, ANSYS Workbench\footnote{www.ansys.com/products/ansys-workbench} was used with ANSYS APDL command snippets to assign material properties as a function of the mesh element coordinates. This FE deformation data was then embedded into a set of synthetic images using numerical image deformation. This process deforms a selected reference image (of user-defined speckle pattern) using FE deformation data, as detailed in \citep{rossi2015eff}. The resulting images are the synthetic equivalent of those obtained experimentally using optical measurements, however the kinematic fields they encode are known and controlled. Grey level noise can also be added to mimic that typical of experimental images. These deformed images were then processed using DIC to obtain strain fields for each load step. In this work, MatchID\footnote{www.matchid.eu} software was used to perform image deformation and DIC. Finally, the proposed algorithm, programmed in MATLAB\footnote{https://matlab.mathworks.com}, was used to perform the parameterisation and identification of constitutive parameters.

This toolchain enables direct comparison of the defined and identified parameter maps, and the errors introduced at each stage to be investigated.

\subsection{Numerical data used to demonstrate the methodology}\label{sec:numerical_data}

\begin{figure}
	\centering
	{\includegraphics[width=0.9\textwidth]{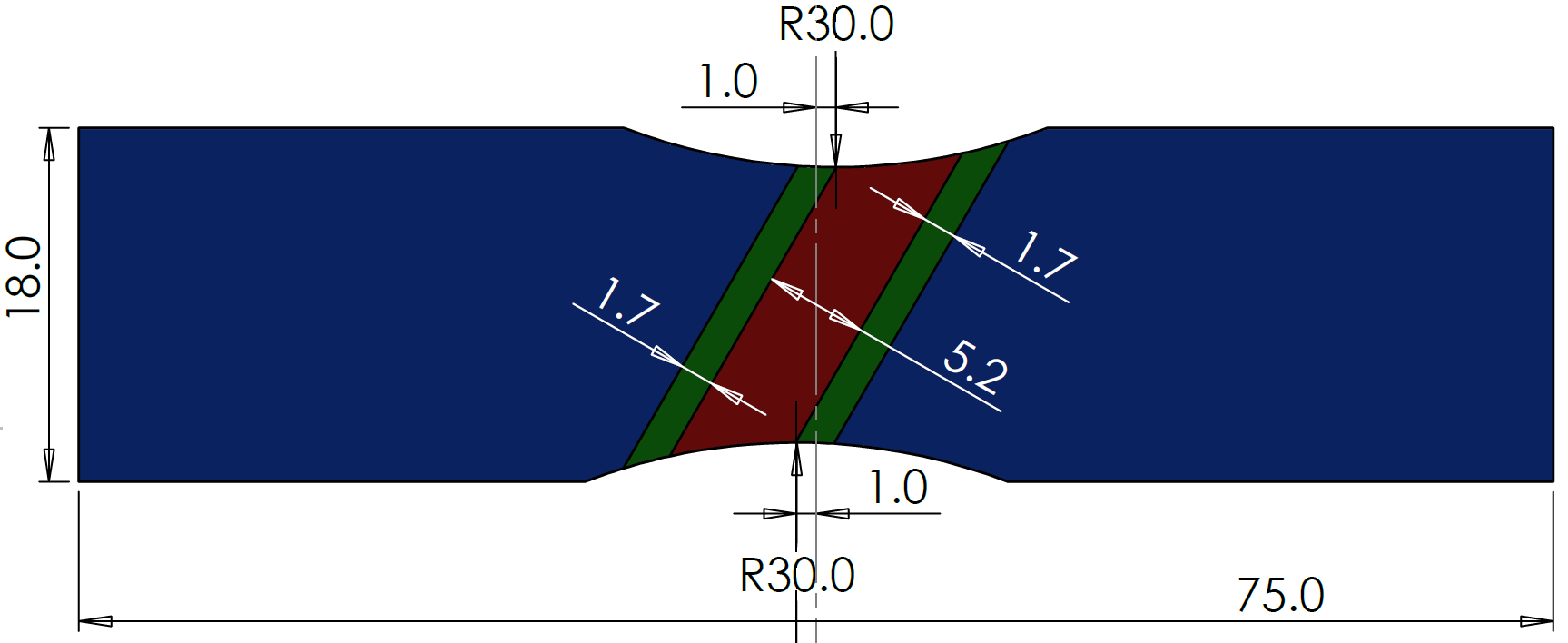}}
	\caption{Schematic of numerical notched butt-weld geometry with dimensions (in mm). Thickness is 1.8~mm.}
	\label{notched_butt_weld_geometry}
\end{figure}
The notched, off-axis, butt-weld geometry shown in \cref{notched_butt_weld_geometry} is used for the demonstrative example. The weld region and heat affected zones have a width of 5.2~mm and 1.7~mm respectively. The off-axis angle is 60 degrees from the specimen longitudinal axis, and two offset notches penetrate the specimen 2~mm on either side. 

\begin{figure}
	\centering
	\includegraphics[width=0.8\textwidth]{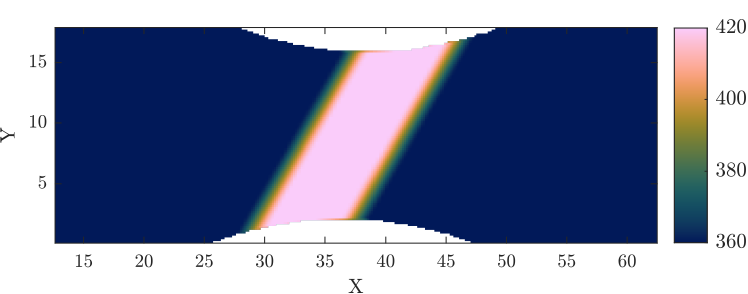}
	\caption{Plot of the assigned yield strength [MPa]}
	\label{fig:notched_weld_yield_strength_map}
\end{figure}

An isotropic von Mises elastoplastic model with linear hardening is used. The elastic properties are assumed known, with a Young’s modulus of 190~GPa and a Poisson’s ratio of 0.28. The yield strength is assigned a value of 360~MPa in the base metal and 420~MPa in the weld metal, with a linear variation throughout the heat affected zones (see \cref{fig:notched_weld_yield_strength_map}). The hardening modulus is assigned a homogeneous value of 3700~MPa. Although it is feasible to apply the developed methodology to a spatially-varying hardening modulus, the plastic response is less influenced by the hardening modulus, so for simplicity it was decided to start with spatial variation of the yield strength only. \cref{sec:conclusion} discusses how future work could incorporate heterogeneous identification of multiple parameters.

On the left-hand edge of the specimen, the horizontal displacement was constrained for all nodes ($u_x^{\textrm{LHS}}=0$). To suppress rigid-body motion, the vertical displacement of a single node at mid-height on the left-hand edge was also fixed ($u_y^{\textrm{LHS-mid}}=0$). On the right-hand edge, a prescribed horizontal displacement, $\delta$, was applied ($u_x^{\textrm{RHS}}=\delta$), and vertical displacements were constrained, $u_y^\textrm{{RHS}}=0$). In this example, a displacement of 1.25~mm was applied (resulting in a force of 12.9~kN) to ensure a good spread of plasticity throughout the specimen. The resulting data was cropped to a region of interest, 50~mm in length, spanning x-coordinates 12.5 and 62.5.

\begin{figure}[h]
	\centering
	\begin{subfigure}[t]{0.49\textwidth}
		\centering
		\includegraphics[width=\textwidth]{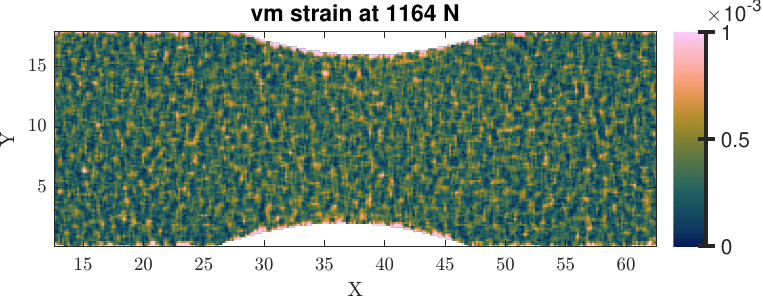}
		\caption{}
		\label{subfig:notched_weld_strain_step_1}
	\end{subfigure}
	\hfill
	\begin{subfigure}[t]{0.49\textwidth}
		\centering
		\includegraphics[width=\textwidth]{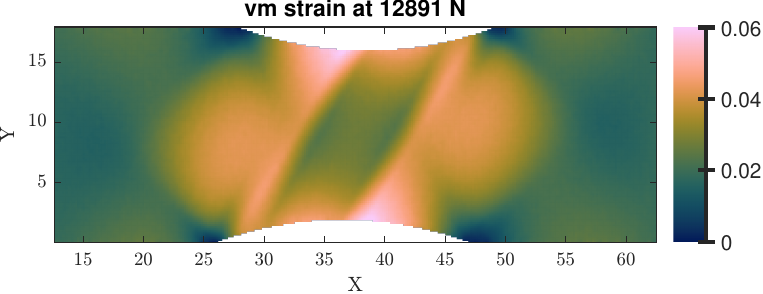}
		\caption{}
		\label{subfig:notched_weld_strain_step_end}
	\end{subfigure}
	\caption[Equivalent strain at first and final load step.]{ \subref{subfig:notched_weld_strain_step_1} Von Mises equivalent strain at first load step. \subref{subfig:notched_weld_strain_step_end} Von Mises equivalent strain at final load step. }
	\label{fig:notched_weld_strain_fields}
\end{figure}

The image deformation toolchain described above was then used with the parameters shown in \cref{table:notched_weld_imDef_DIC} to generate synthetic strain fields representative of a typical experiment. Grey level noise was added to the images, leading to a longitudinal strain noise floor of 140~microstrain. The von Mises equivalent strain fields for the first and final load steps are shown in \cref{fig:notched_weld_strain_fields}. 

\begin{table}[h]
	\centering
	\small
	\caption{Key parameters selected for image deformation and digital image correlation.}
	\label{table:notched_weld_imDef_DIC}
	\begin{tblr}{
			colspec={Q[r,m] Q[l,m]}, 
			width=\textwidth,
			row{1,6,13}={font=\bfseries\boldmath},  
			hline{1,16}={1.2pt},        
			hline{2,6,7,13,14}={0.5pt},    
		}
		\SetCell[r=1,c=2]{l} Image deformation settings \\
		Speckle pattern & 4096 $\times$ 2160 pixels  \\ 
		Average speckle size & 3.5 pixels  \\ 
		Interpolation & Global bicubic spline  \\
		Noise & 3.84 grey levels (1.5\%) \\
		\SetCell[r=1,c=2]{l} Digital image correlation settings \\
		Correlation criterion & Approximated sum of square differences, ASSD  \\ 
		Subset weight & Gaussian  \\
		Shape function & Quadratic \\
		Interpolation & Local bicubic splines\\
		Subset size & 27\\
		Step size & 12\\
		\SetCell[r=1,c=2]{l} Strain calculation settings \\
		Strain window & 5\\
		Strain interpolation & Linear quadrilateral (Q4)\\
	\end{tblr}
\end{table}
%

\subsection{Identification overview}
The methodology comprises a preliminary and a refined identification phase, with each requiring a parameterisation scheme, cost function and optimiser to be defined. 

Each constitutive parameter is designated as: known, homogeneous or heterogeneous. If known, the user provides values of the constitutive parameter for each data point and these remain unchanged throughout identification. If assigned homogeneous, a single value is to be identified for all data points of this parameter type. If assigned heterogeneous, some combination of the parameterisation schemes discussed in \cref{sec:parameterisation_of_const_param} is used. Hence, the constitutive parameters at each data point are fully defined by the degrees of freedom (DOFs) of the specified parameterisation. 

For a given set of DOFs, the corresponding parameter maps can be combined with the known strain fields to reconstruct the stress fields. A specified cost function is then used to assess stress equilibrium throughout the specimen. The DOFs can then be iterated using the defined optimisation scheme until the solution converges to the parameter maps which minimise error in the stress equilibrium. In this work, stress reconstruction was performed using the radial return algorithm developed by \citeauthor{marek2017sen} as described in \citep{marek2017sen}, based on the formulation presented in \citep{e.a.desouzaneto2008com}. Ongoing work is investigating alternative approaches, such as the NEML2 package \citep{hu2025nem}, to further improve computational performance.  

\subsection{Phase 1: Preliminary identification}
The goal of the preliminary identification is to provide an initialisation for the refined identification phase. The constitutive properties are parameterised as homogeneous, with a uniform value across the specimen. The sensitivity-based virtual fields are used with the cost function defined in \cref{eq:sbvf_cost_function}. As the total number of unknowns is small, the cost function is straightforward and a gradient-based optimiser can be used. In this example, an initial value of 320~MPa and 3000~MPa is assumed for the yield strength and hardening modulus respectively. The Levenberg–Marquardt gradient-based optimisation identified values of 385~MPa and 3299~MPa in seven iterations (30 seconds). 

\subsection{Phase 2: Refined identification}\label{sec:phase_2_inter_ident}
The refined identification utilises the EGI and FRE metrics along with basis function parameterisation.

\subsubsection{Cost function}
As discussed in \cref{sec:stress_equil_metrics}, the EGI and FRE metrics play complementary roles in assessing stress equilibrium. The EGI evaluates how well the distribution of stress within an isolated window satisfies equilibrium, whereas the FRE quantifies how well the reconstructed longitudinal force balances the applied force for each slice. For optimisation these terms are combined to return a single scalar, $\globalCost$, using \cref{eq:global_cost},
\begin{equation}
\globalCost
=
(1-\mixWeight)\,\egiCost
+
\mixWeight\,\freCost .
\label{eq:global_cost}
\end{equation}
$\mixWeight$ is a user-defined weight between 0 and 1 that controls the relative contribution of each term. In this work $\mixWeight=0.1$ was selected, as preliminary investigation indicated it performed well for the example problem. However, future work should perform sensitivity studies across a wider range of applications in order to evaluate its robustness.

The EGI contribution, $\egiCost$ (\cref{eq:phi_egi}), aggregates the force-weighted  spatial-temporal RMS values (defined in \cref{eq:EGI_weighted_spatiotemporal_rms}) computed for a set of window sizes indexed by $k = 1,\ldots,\numWinSizes$. A window-size weight, $\winWeight_k$, is used so that the contribution from smaller windows, which are more sensitive to local discrepancies in equilibrium, does not drown out the contribution of larger windows that enforce equilibrium at a more global scale. $\winWeight_k$ is proportional to the mean window side length, $\winLength_k$, and normalised so $\sum_{k}\winWeight_k = 1$. $\egiScale^{k}$ is discussed below.
\begin{equation}
\egiCost
=
\sum_{k=1}^{\numWinSizes}
\frac{\winWeight_k}{\egiScale^{k}}\,
\mathrm{EGI}^{\,k}_{\mathrm{weighted\_spatial\text{-}temporal\_RMS}} .
\label{eq:phi_egi}
\end{equation}
\begin{equation}
\winWeight_k = \frac{\winLength_k}{\sum_{j=1}^{\numWinSizes}\winLength_j},
\qquad
\sum_{k=1}^{\numWinSizes}\winWeight_k = 1 .
\label{eq:win_weight}
\end{equation}
The FRE contribution, $\freCost$ (\cref{eq:phi_fre}) uses the force-weighted spatial-temporal RMS defined in \cref{eq:FRE_weighted_spatiotemporal_rms}. 
\begin{equation}
\freCost
=
\frac{1}{\freScale}\,
\mathrm{FRE}_{\mathrm{weighted\_spatial\text{-}temporal\_RMS}},
\label{eq:phi_fre}
\end{equation}
The factors $\egiScale^{k}$ and $\freScale$ scale the contributions of the EGI and FRE terms so they are dimensionless and of a comparable order of magnitude.  In this work, $\egiScale^{k}$ and $\freScale$ are defined using a reference stress field, $\stressRef$, to provide an order of magnitude scaling consistent with the noise level in the data. Here, $\stressRef$ is set to the stress field identified in Phase 1. Although this stress field is not assumed to satisfy equilibrium at this stage, it was found adequate to prevent either term overwhelming the other. Alternatively, a user-defined region assumed to be homogeneous can be identified in a piece-wise manner, and the resulting stress used as the reference stress. This was trialled and produced comparable scaling, however, the Phase 1 result is used here for simplicity. 
\begin{equation}
\egiScale^{k}
=
\left.
\mathrm{EGI}^{\,k}_{\mathrm{weighted\_spatial\text{-}temporal\_RMS}}
\right|_{\stress=\stressRef} 
\label{eq:egi_weight}
\end{equation}
\begin{equation}
\freScale
=
\left.
\mathrm{FRE}_{\mathrm{weighted\_spatial\text{-}temporal\_RMS}}
\right|_{\stress=\stressRef} 
\label{eq:fre_weight}
\end{equation}

Future work should investigate the sensitivity of the identification to $\mixWeight$ and the scaling factors $\egiScale^{k}$ and $\freScale$, with the aim of reducing user-defined choices. 

\subsubsection{EGI windows}\label{sec:defining_egi_windows}
The optimal EGI window size depends on the length scale of heterogeneity in the unknown constitutive parameter maps and measurement noise. Larger windows smooth the data and reduce noise sensitivity, whereas smaller windows provide greater spatial resolution, but are more susceptible to noise. 

As the EGI formulation excludes external virtual work, the EGI windows cannot intersect traction boundaries. This introduces a border of missing data at traction boundaries equal to half the window size. Hence, overly large windows may result in significant loss of EGI data at traction edges. No such border exists at traction-free boundaries as the EGI window can cross free edges without issue. Tackling this problem of missing data using interpolation or adaptive window sizing is left for future work.

The EGI formulation using a nine-noded mesh requires a minimum window size of three data points in each dimension. In practice, the minimum window size is limited by the spatial resolution of the EGI noise floor. Below a certain window size, the equilibrium error due to incorrect parameter distribution will be drowned out by the error resulting from the strain noise. 

\begin{figure}
	\centering
	\includegraphics[width=0.8\textwidth, trim=0mm 0mm 0mm 0mm,clip]{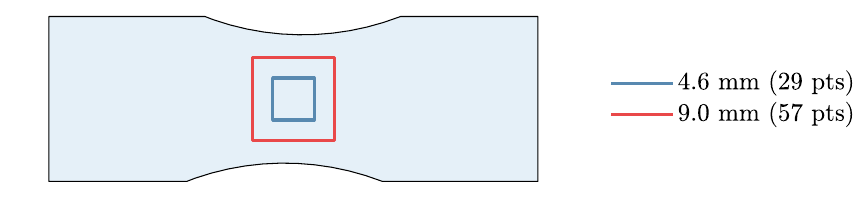}
	\caption[Selected EGI windows for off-axis butt weld identification.]{Plot showing selected EGI window sizes}
	\label{fig:egi_selected_windows}
\end{figure}

For simplicity, two fixed window sizes were used throughout identification in this work. The larger window side length was set to 50\% of the smaller specimen dimension (9~mm for a specimen height of 18~mm) and the smaller window to  25\% (4.5~mm), corresponding to 57 and 29 data points respectively. These windows are visualised in \cref{fig:egi_selected_windows}.  This choice proved sufficient to enforce equilibrium across relevant length-scales. Additional windows could be used, but windows that are too small (noise dominated) or too large (boundary limited) contribute little sensitivity to the cost function while increasing computation cost. Future work should investigate identification sensitivity to the number and size of EGI windows for different problems. A more automated approach could also be developed to define (and potentially update) window size for optimum identification.

\subsubsection{Spatial parameterisation}\label{sec:init_param_refined_ident}
The refined identification phase utilises a homogeneous region with one or more basis functions to parameterise any heterogeneous constitutive parameters. The homogeneous region includes all specimen data points and acts as a `floor' from which the basis functions can add or detract. This is well-suited to the application of welded joints, for which the base metal(s) tend to be largely homogeneous with the weld region being under- or over-matched.

\begin{figure}[h]
	\centering
	\includegraphics[width=\textwidth]{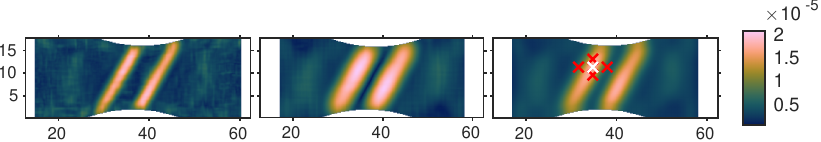}
	\caption[Equilibrium-gap maps and candidate centres used to initialise the new basis-function location.]{
	Weighted temporal RMS maps of EGI for 29x29 window (left) and 57x57 window (middle). The smoothed combined map (right) is then used to initialise centres. The white cross indicates the centre located at the max combined EGI value. The four red crosses are at the perturbed centres of the multi-start candidates.}
	\label{fig:combined_egi}
\end{figure}

For simplicity, a single basis function is added to the parameterisation after each global iteration. The force-weighted temporal RMS of the equilibrium gap indicator, $\mathrm{EGI}_{\mathrm{weighted\_spatial\text{-}temporal\_RMS}}$ (\cref{eq:EGI_weighted_temporal_rms}), provides a spatial map of stress equilibrium discrepancies. For each EGI window size, these maps are computed and then combined using the window-length weights $\winWeight_k$ from \cref{eq:win_weight}. This produces a combined equilibrium gap map consistent with the multi-window weighting used in the EGI contribution to the cost function \cref{eq:phi_egi}. This combined map is smoothed using a moving-window mean filter (3~mm square in this example) to remove sensitivity to local noise, and the location of the maximum value is used as an initial centre of the new basis function. This initial guess is then refined using a multi-start procedure in which five candidates are evaluated - the initial centre and four perturbed centres (perturbed by 10~\% of the coordinate range in each direction). Each candidate is refined using the cost function and 10 pattern search iterations. The best refined candidate is then used to initialise the coordinates of the new RBF in the main optimisation (described below). This process is visualised in \cref{fig:combined_egi}.

\subsubsection{Optimisation}
For a given parameterisation, the degrees of freedom (DOF) fully define the constitutive parameters for every data point. Combining these parameter maps with the measured strain fields returns the stress fields, which can then be evaluated using the cost function defined in \cref{eq:global_cost}. Thus the DOFs can be optimised to minimise the cost function to obtain the parameter maps that best satisfy stress equilibrium. 

Each identification iteration is split into two stages. First, the floor term is fixed (initialised from the previous iteration) and the basis function DOF (x-coordinate, y-coordinate, height and variance) are optimised. Then, once converged, the x and y coordinates of the basis functions are fixed, and the floor is optimised alongside the height and variance terms. Although all DOF could be optimised simultaneously, this approach seemed more robust - particularly for problems in which large variances may result in the basis functions competing with the floor term. 

The identification procedure is terminated when updating the parameterisation (adding a basis function) reduces the objective by less than a user-defined percentage (5~\% in this work).

When multiple basis functions are present many local minima may exist, so a global optimisation approach is required. 
In this work, MATLAB's pattern search algorithm was used for optimisation (see \citep{matlab2024glo} for more information). The pattern search algorithm evaluates the cost function for a set of points, called a mesh or pattern, around the current solution. In this work, the pattern evaluates a lower and higher value for each degree of freedom, resulting in two cost function evaluations per DOF per iteration. If any candidate solution improves on the current solution, the current solution is updated and the pattern expands. If no better solution is found, the pattern contracts. This process is repeated until the solution converges. The expansion and contraction factors can be modified, however, the default values of 0.5 and 2 provide a good balance of exploration and exploitation. This is somewhat analogous to a gradient-based optimiser that halves or doubles its step size depending on the success of each evaluation. The pattern search is a powerful optimiser with minimal user-tuning required. It also works well in combination with the multi-start approach shown in \cref{fig:combined_egi}. However, other optimisers such as genetic algorithms were also tried successfully, and future work could explore others such as simulated annealing or particle swarm algorithms. 

\subsection{Identification results using univariate Gaussian functions}
\cref{tab:univariate_results_summary} presents the identified yield strength and hardening modulus at each iteration. The yield strength plots are masked such that data points with an identified plastic strain below 0.5~\% are semi-transparent, indicating the identified value is unreliable due to insufficient plastic strain. This is limited to small regions at the edge of the notch in this case. The identification proceeds as intended, automatically converging toward the target parameter map without \textit{a priori} information on the distribution of properties. The identification converges when four basis functions are used (adding a fifth yields little improvement). 

%
\newlength{\imageHeightOne}
\setlength{\imageHeightOne}{2.6cm}
\newcommand{\resTabImg}[1]{%
  \raisebox{-.5\height}{%
    \includegraphics[
      width=\linewidth,
      height=\imageHeightOne,
      keepaspectratio,
    ]{#1}%
  }%
}
\begin{table*}[h!]
\centering
\small
\caption{Identified yield strength, $\stress_Y$, and hardening modulus, $H$, at each iteration using univariate functions.}
\label{tab:univariate_results_summary}
\begin{tblr}{
  width=\textwidth,
  colspec = {
    X[0.1,c,m]  
    X[0.5,c,m]  
    X[0.15,c,m] 
    X[0.15,c,m] 
  },
  rowsep = 3pt,
  colsep = 4pt,
  hline{1,Z}={1.2pt},
  hline{3}={0.5pt},       
  column{1} = {font=\bfseries},
  row{1-2} = {font=\bfseries},
  row{4-8} = {ht=\imageHeightOne}, 
}
\SetCell[r=2]{c}\rotatebox{90}{\strut Iter} &
\SetCell[c=2]{c} Identified [MPa] & &
Duration \\
\cmidrule[wd=0.2pt]{2-3}
& $\stress_{y}$ & $H$  & [Minutes] \\
1 & 385 & 3299 & 0.6 \\
2 & \resTabImg{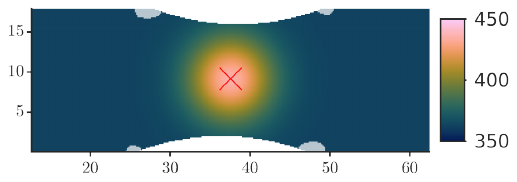} & 2174 & 14.2 \\
3 & \resTabImg{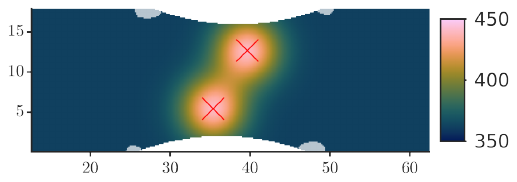} & 3009 & 40.8 \\
4 & \resTabImg{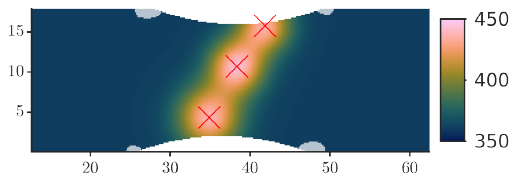} & 3157 & 81.4 \\
5 & \resTabImg{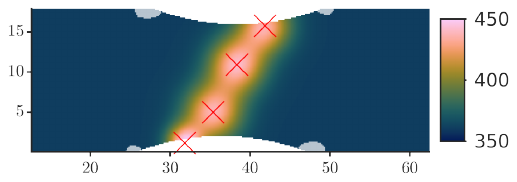} & 3220 & 70.8 \\
6 & \resTabImg{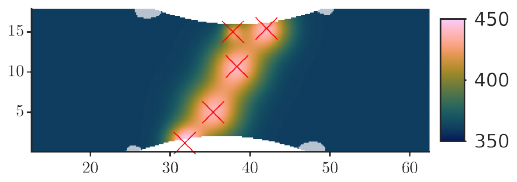} & 3234 & 72.6 \\
\end{tblr}
\end{table*}

\begin{figure}[h]
	\centering
	\includegraphics[width=0.8\textwidth]{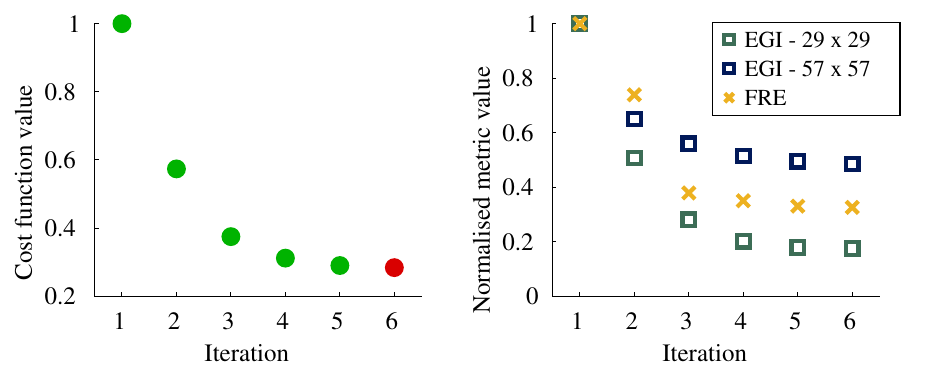}
	\caption[Convergence plot for the refined identification of the example data using univariate basis functions.]{Convergence plot for the identification using univariate basis functions.}
	\label{fig:univariate_convergence_plot}
\end{figure}

The use of synthetic data, for which the target parameters are known, enables the errors at each data point to be quantified. The absolute percentage error for the hardening modulus and yield strength is presented in \cref{tab:univariate_error_metrics}. The final hardening modulus value is 3234~MPa, approximately 13\% lower than the target 3700~MPa. The identified yield strength has a value of 360.7~MPa in the base metal (0.2\% higher than target) with a maximum absolute percentage error of approximately 6~\% around the weld metal. The remaining error is largely attributed to the parameterisation struggling to capture the gradient across the HAZ and the uniform distribution throughout weld using Gaussian functions. This was confirmed by fitting four Gaussians directly to the target map, yielding a similar distribution and magnitude of error.

\cref{tab:univariate_error_metrics} also shows the weighted temporal RMS of the EGI and FRE metrics for the identified stress field at each iteration. The metrics do a good job at identifying regions in which the yield strength is incorrect. As expected the smaller window provides better spatial resolution but is noisier. The FRE serves to keep the overall magnitudes of force balance reasonable, and after iteration 3 the reconstructed longitudinal stress is within 2\% of the applied force throughout the majority of the specimen.

\clearpage
\newgeometry{left=15mm,right=15mm,top=20mm,bottom=20mm} 
\begin{landscape}
\newlength{\imageHeight}
\setlength{\imageHeight}{2.15cm} 
\newcommand{\tblimg}[1]{%
\raisebox{-.5\height}{%
  \includegraphics[
    width=\linewidth,
    height=\imageHeight,
    keepaspectratio,
  ]{#1}}%
}
\newcommand{\tblcbar}[1]{%
  \includegraphics[
    width=\linewidth,
    height=\imageHeight,
    keepaspectratio,
  ]{#1}%
}
\begin{table}[p]
\caption{Absolute percentage errors of the identified hardening modulus, $H$, and yield strength, $\stress_Y$, together with the corresponding weighted temporal RMS maps for the EGI (\cref{eq:EGI_weighted_temporal_rms}) and FRE (\cref{{eq:FRE_weighted_temporal_rms}}) metrics. Where applicable, a shared colour bar is provided at the top of each column. The EGI and FRE colour scales are logarithmic.}
\label{tab:univariate_error_metrics}
\centering
\begin{tblr}{
  width=\linewidth,
  colspec = {
    X[0.3,c,m]
    X[0.4,c,m]
    X[2.4,c,m]
    X[2.4,c,m]
    X[2.4,c,m]
    X[2.4,c,m]
  },
  rowsep = 3pt,
  colsep = 2pt,
  hline{1,10}={1.2pt},
  hline{3}={0.5pt},
  column{1} = {font=\bfseries},
  row{1} = {font=\bfseries},
  row{3} = {ht=\imageHeight},
}
\SetCell[r=2]{c}\rotatebox{90}{\strut Iter.} & \SetCell[c=2]{c} Absolute percentage error & & \SetCell[c=3]{c} Weighted temporal RMS & & \\
\cmidrule[wd=0.2pt,r=-1.2]{2-3}
\cmidrule[wd=0.2pt,l=-1.2]{4-6}
& H & $\stress_Y$ & EGI 29$\times$29 & EGI 57$\times$57 & FRE \\
& & \tblcbar{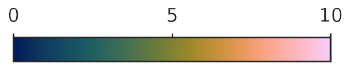} & \tblcbar{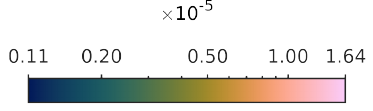} & \tblcbar{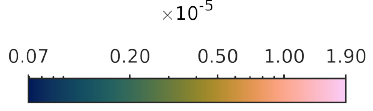} & \tblcbar{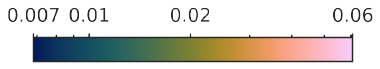} \\
1 & 10.8 & \tblimg{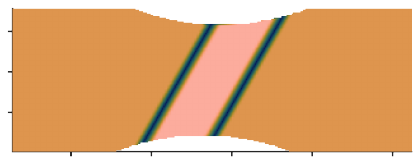} & \tblimg{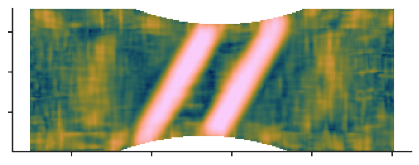} & \tblimg{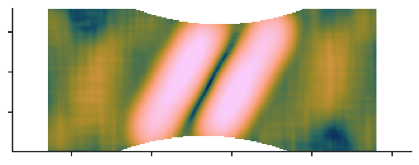} & \tblimg{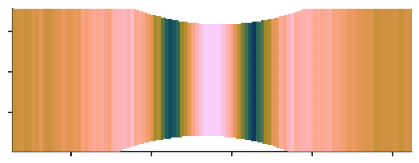} \\
2 & 41.2 & \tblimg{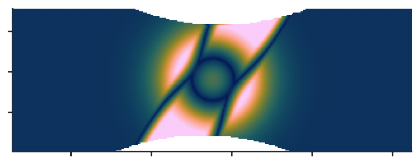} & \tblimg{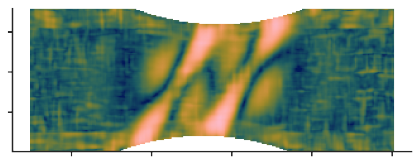} & \tblimg{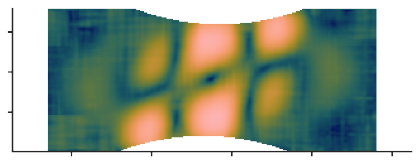} & \tblimg{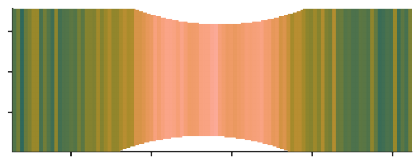} \\
3 & 18.7 & \tblimg{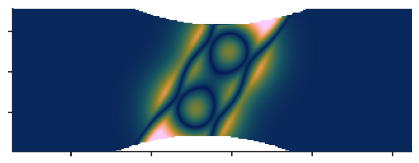} & \tblimg{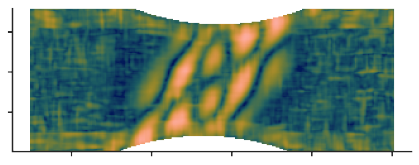} & \tblimg{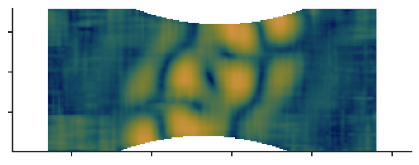} & \tblimg{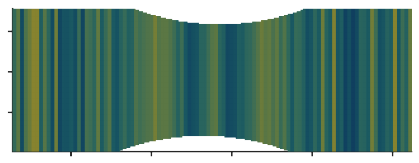} \\
4 & 14.7 & \tblimg{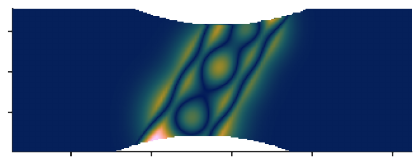} & \tblimg{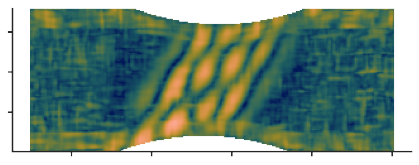} & \tblimg{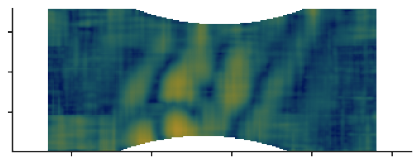} & \tblimg{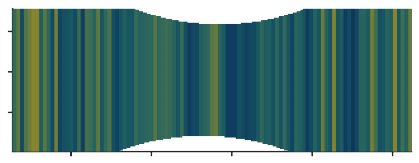} \\
5 & 13 & \tblimg{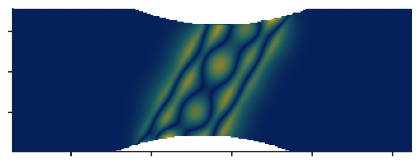} & \tblimg{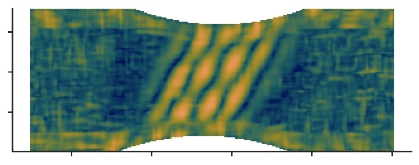} & \tblimg{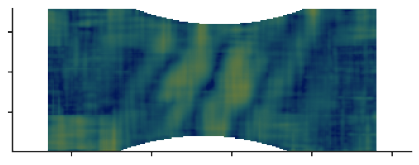} & \tblimg{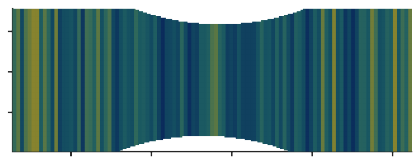} \\
6 & 12.6 & \tblimg{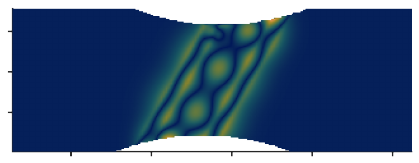} & \tblimg{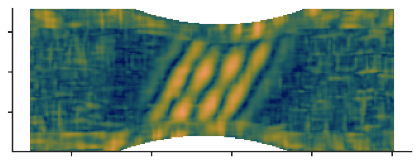} & \tblimg{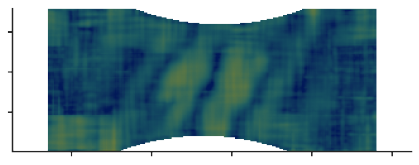} & \tblimg{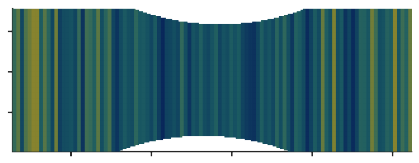} \\
\end{tblr}
\end{table}
\end{landscape}
\restoregeometry
\clearpage

\subsection{Identification results using bivariate Gaussian functions}\label{sec:recommended_approach_example}
When bivariate Gaussians are used in the parameterisation, the identification converges with a single bivariate Gaussian function as shown in \cref{tab:bivariate_results_summary}.

The final hardening modulus value is 3383~MPa, approximately 9\% lower than the target 3700~MPa. The identified yield strength has a value of 359.7~MPa in the base metal (0.1\% lower than target) with a maximum absolute percentage error of approximately 4~\% around the weld metal. 

Adding a second basis function serves to better capture the gradient throughout the left hand HAZ region, and this improvement can be seen in the EGI 29x29 result in \cref{tab:bivariate_error_metrics}. However, the improvement in the parameterisation is insufficient to meet the 5~\% convergence criteria.

\newlength{\imageHeightOnebivariate}
\setlength{\imageHeightOnebivariate}{2.6cm}
\newcommand{\resTabImgbivariate}[1]{%
  \raisebox{-.5\height}{%
    \includegraphics[
      width=\linewidth,
      height=\imageHeightOnebivariate,
      keepaspectratio,
    ]{#1}%
  }%
}
\begin{table*}[h]
\centering
\small
\caption{Identified yield strength, $\stress_Y$, and hardening modulus, $H$, at each iteration using bivariate functions.}
\label{tab:bivariate_results_summary}
\begin{tblr}{
  width=\textwidth,
  colspec = {
    X[0.1,c,m]  
    X[0.5,c,m]  
    X[0.15,c,m] 
    X[0.15,c,m] 
  },
  rowsep = 3pt,
  colsep = 4pt,
  hline{1,Z}={1.2pt},
  hline{3}={0.5pt},        
  column{1} = {font=\bfseries},
  row{1-2} = {font=\bfseries},
  row{4-5} = {ht=\imageHeightOnebivariate},
}
\SetCell[r=2]{c}\rotatebox{90}{\strut Iter} &
\SetCell[c=2]{c} Identified [MPa] & &
Duration \\
\cmidrule[wd=0.2pt]{2-3}
& $\stress_{y}$ & $H$  & [Minutes] \\
1 & 385 & 3299 & 0.7 \\
2 & \resTabImgbivariate{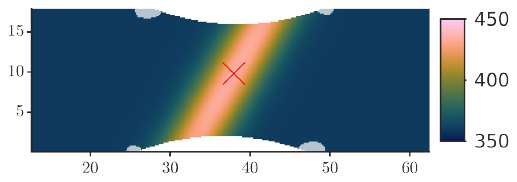} & 3327 & 32.4 \\
3 & \resTabImgbivariate{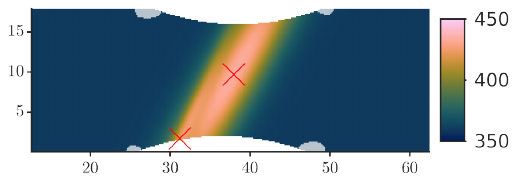} & 3383 & 39.8 \\
\end{tblr}
\end{table*}

%
\clearpage
\newgeometry{left=15mm,right=15mm,top=20mm,bottom=20mm} 
\begin{landscape}
\newlength{\imageHeightbivariate}
\setlength{\imageHeightbivariate}{2.15cm}
\newcommand{\tblimgbivariate}[1]{%
\raisebox{-.5\height}{%
  \includegraphics[
    width=\linewidth,
    height=\imageHeightbivariate,
    keepaspectratio,
  ]{#1}}%
}
\newcommand{\tblcbarbivariate}[1]{%
  \includegraphics[
    width=\linewidth,
    height=\imageHeightbivariate,
    keepaspectratio,
  ]{#1}%
}
\begin{table}[p]
\caption{Absolute percentage errors of the identified hardening modulus, $H$, and yield strength, $\stress_Y$, together with the corresponding weighted temporal RMS maps for the EGI (\cref{eq:EGI_weighted_temporal_rms}) and FRE (\cref{{eq:FRE_weighted_temporal_rms}}) metrics. Where applicable, a shared colour bar is provided at the top of each column. The EGI and FRE colour scales are logarithmic.}
\label{tab:bivariate_error_metrics}
\centering
\begin{tblr}{
  width=\linewidth,
  colspec = {
    X[0.3,c,m]
    X[0.4,c,m]
    X[2.4,c,m]
    X[2.4,c,m]
    X[2.4,c,m]
    X[2.4,c,m]
  },
  rowsep = 3pt,
  colsep = 2pt,
  hline{1,7}={1.2pt},
  hline{3}={0.5pt},
  column{1} = {font=\bfseries},
  row{1} = {font=\bfseries},
  row{3} = {ht=\imageHeightbivariate},
}
\SetCell[r=2]{c}\rotatebox{90}{\strut Iter.} & \SetCell[c=2]{c} Absolute percentage error & & \SetCell[c=3]{c} Weighted temporal RMS & & \\
\cmidrule[wd=0.2pt,r=-1.2]{2-3}
\cmidrule[wd=0.2pt,l=-1.2]{4-6}
& H & $\stress_Y$ & EGI 29$\times$29 & EGI 57$\times$57 & FRE \\
& & \tblcbarbivariate{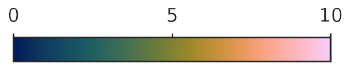} & \tblcbarbivariate{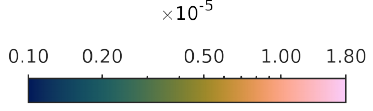} & \tblcbarbivariate{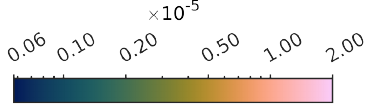} & \tblcbarbivariate{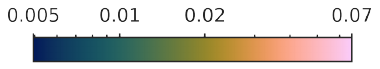} \\
1 & 10.8 & \tblimgbivariate{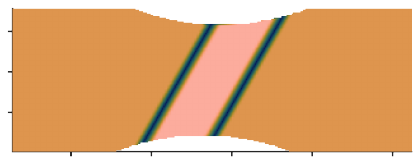} & \tblimgbivariate{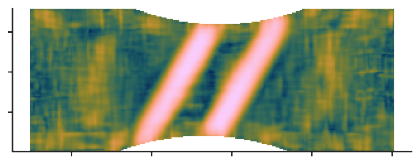} & \tblimgbivariate{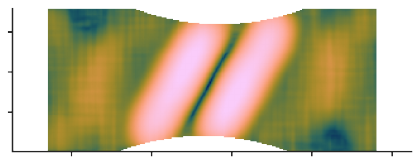} & \tblimgbivariate{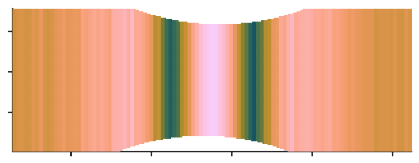} \\
2 & 10.1 & \tblimgbivariate{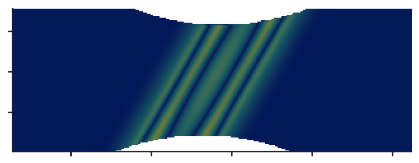} & \tblimgbivariate{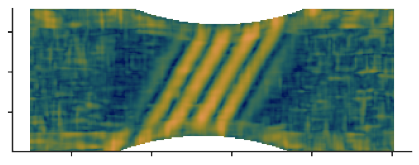} & \tblimgbivariate{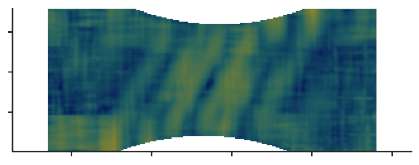} & \tblimgbivariate{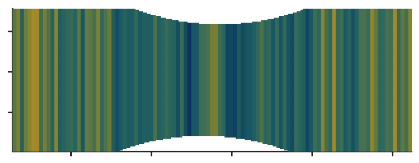} \\
3 & 8.6 & \tblimgbivariate{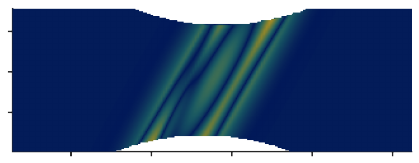} & \tblimgbivariate{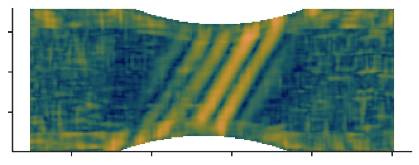} & \tblimgbivariate{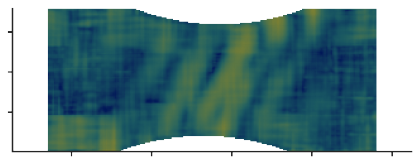} & \tblimgbivariate{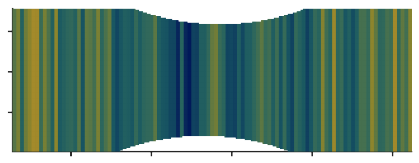} \\
\end{tblr}
\end{table}
\end{landscape}
\restoregeometry
\clearpage
\section{Discussion and future work}\label{sec:discussion}
The results demonstrate that the proposed methodology identifies the target constitutive parameters without requiring \textit{a priori} information about their spatial distribution. Automated refinement of the spatial parameterisation is enabled by the equilibrium metrics, with the EGI highlighting local equilibrium discrepancies and the FRE promoting global force balance throughout the specimen. 

As the identification converges, each metric approaches its noise floor, with the smaller EGI window remaining informative for longer as remaining equilibrium discrepencies become more local. Further refinement could be sought by lowering the convergence threshold (\textit{e.g.} from 5\% to 1\% improvement), introducing another smaller EGI window and/or increasing the weight assigned to the smaller window. Preliminary investigations in which smaller windows were added each time the identification converged showed promise, but future work should systematically explore adaptive strategies to select the quantity, number and weighting of EGI windows as the identification progresses. 

Ultimately, the achievable results will be limited by the noise floor and limitations of the parameterisation. In this example, the bivariate parameterisation achieved lower error in less time than the univariate form (73 minutes versus 280 minutes). This was expected as the bivariate function is more suitable for reconstructing the elliptical-like distribution of properties throughout the off-axis, butt-weld geometry. More generally, the optimal parameterisation depends on the application. For example, the univariate Gaussians are likely more efficient at capturing spatial variation characterised by localised hotspots (such as spot welds). Future work should explore additional parameterisation strategies including basis functions capable of higher spatial frequencies or the use of splines.

The identified hardening modulus was consistently lower than the target value. This is likely caused by limited plastic strain (spatially and temporally), and might benefit from alternative weighting of the loadsteps (\textit{e.g.} higher weight assigned to steps with larger plastic strain or equilibrium error). Furthermore, the hardening modulus was identified as homogeneous in this work. To enable heterogeneous identification, a sensible starting point may be to assume the distribution of the hardening modulus is the same as that for the yield strength. Hence, the converged parameterisation of the identified yield strength could be reused to identify the hardening modulus, keeping the basis function positions fixed and identifying only their heights and widths. Research is required to explore the  validity of this approach. More broadly, the methodology can be extended to alternative constitutive models depending on the required application.

\begin{figure}
	\centering
	\includegraphics[width=0.8\textwidth, trim=0mm 0mm 0mm 0mm,clip]{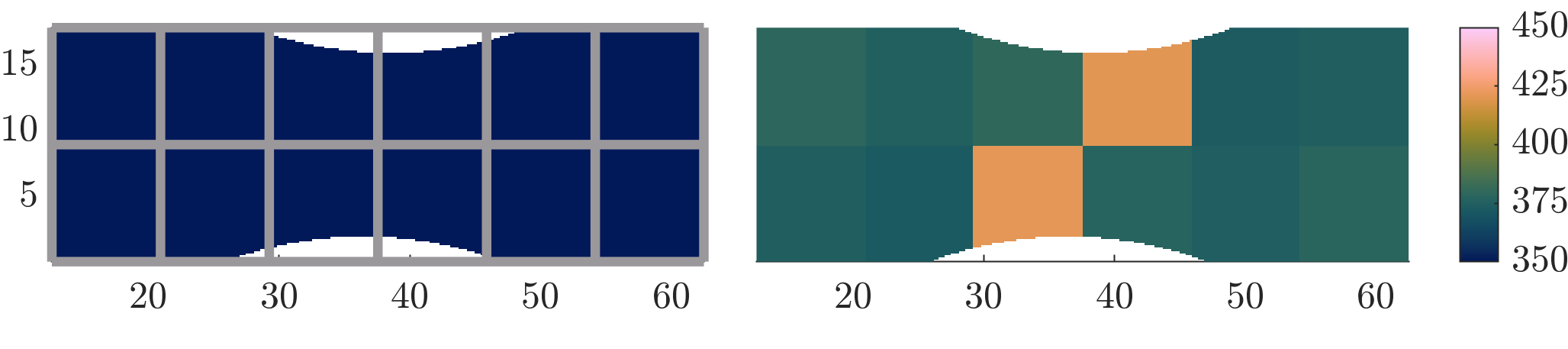}
	\caption{Initial and identified result of yield strength using a zero-order 2 x 8 mesh.}
	\label{fig:mesh_result}
\end{figure}

Despite efforts to reduce user-defined choices and minimise sensitivity to those that remain, more rigorous sensitivity analysis is required. As the methodology matures, there is significant scope for adaptation and extension. The proposed methodology is intended to demonstrate how spatial parameterisation, stress equilibrium metrics and optimisation can be combined in a flexible framework. For example, a coarse zero-order mesh can be used during the preliminary identification from which basis functions could be fitted to accelerate the refinement process (see \cref{fig:mesh_result}). 

Computational times are reasonable (approximately one to five hours on an Intel® Core™ i7-9700 CPU at 3 GHz with 32~GB of RAM, depending on the selected parameterisation), however, substantial improvements are expected through refactoring and parallelisation. The pattern-search optimiser is ideal for multi-core computation, however was not used in parallel here. Furthermore, at this development stage, runtimes are extended due to conservative debugging and output generation. Work is ongoing to refactor the code for imrproved performance, extensibility and dissemination. 

Finally, experimental validation across a range of weld materials, geometries and loading configurations is required to demonstrate credibility of the approach and enable real-world impact on the design and performance assessment of welded joints. It is worth noting that the apparent yield stress identified here includes any contributions from residual stresses. If necessary, non-destructive residual stress measurement techniques could be employed, and the residual stress field subtracted from the stress field identified using the current methodology.

In the longer term, enhanced integration between the numerical and experimental components in the toolchain (\cref{fig:toolchain}) could enable simulation-led testing and uncertainty quantification at each stage of the process. Sensitivity studies could then be used to optimise the geometry, loading configuration, and test set-up to maximise the signal-to-noise ratio of the parameters being identified. Such work would support the broader effort towards probabilistic assessment of component performance under relevant environmental conditions.

\section{Conclusion}\label{sec:conclusion}
This paper has presented the development of a methodology to map the heterogeneous elastoplastic mechanical properties of welded joints. Capitalising on information-rich, full-field optical measurements, this approach extends the virtual fields method (VFM) by introducing automated spatial parameterisation of the constitutive parameters. This reduces reliance on \textit{a priori} information about the mechanical property distribution and enables the novel characterisation of welds with more complex geometries, loading conditions and dissimilar materials.

The methodology was verified using a numerical toolchain that generated synthetic images representative of those captured experimentally, for an off-axis butt weld geometry. The results demonstrate that the proposed approach automatically converged towards the target parameter maps without \textit{a priori} information on the distribution of the properties. The identified yield-strength map had a maximum error within 6\% of the target values with a acceptable runtime. 

These results demonstrate the methodology as a proof of concept, however further work is required to establish robustness and credibility. Future work should prioritise sensitivity analysis, investigation with a broader range of weld configurations and loading conditions, and experimental validation. Ongoing work to refactor the code is also expected to improve computational performance and extensibility.\footnote{These tools currently exist as proof-of-concept algorithms and the refactored code will be made publicly available for collaboration with documentation and examples.} With further development, these tools could support the design, manufacture and validation of complex components - from joins in multi-material components to additively manufactured parts - accelerating fusion power plant delivery and enabling broader industrial impact. 

\section*{CRediT authorship contribution statement}
\noindent \textbf{Robert Hamill:} Investigation, Methodology, Software, Formal analysis, Visualization, Writing -- original draft, Writing -- review \& editing.
\textbf{Allan Harte:} Supervision, Funding acquisition, Writing -- review \& editing.
\textbf{Aleksander Marek:} Conceptualization, Software, Supervision.
\textbf{Fabrice Pierron:} Conceptualization, Supervision, Project administration, Writing -- review \& editing.

\section*{Acknowledgements}
This work was supported by the Engineering and Physical Sciences Research Council (EPSRC) under grant EP/W006839/1. The authors would also like to acknowledge support from the University of Southampton and the United Kingdom Atomic Energy Authority.

\bibliographystyle{elsarticle-num-names}
\bibliography{references} 

\end{document}